\documentclass{emulateapj}

\begin{document}

\title{Disk-Jet Connection in the Radio Galaxy 3C 120}

\author{Ritaban Chatterjee\altaffilmark{1}, Alan P. Marscher\altaffilmark{1}, Svetlana G. Jorstad\altaffilmark{1,2}, 
Alice R. Olmstead\altaffilmark{1}, Ian M. McHardy\altaffilmark{3}, Margo F. Aller \altaffilmark{4}, Hugh D. Aller\altaffilmark{4}, Anne L\"ahteenm\"aki\altaffilmark{5}, Merja Tornikoski\altaffilmark{5}, Talvikki Hovatta\altaffilmark{5}, Kevin Marshall\altaffilmark{6,7}, H. Richard Miller\altaffilmark{6}, Wesley T. Ryle\altaffilmark{6}, Benjamin Chicka\altaffilmark{1}, A. J. Benker\altaffilmark{8,9}, Mark C. Bottorff\altaffilmark{10}, David Brokofsky\altaffilmark{8,11}, Jeffrey S. Campbell\altaffilmark{8}, Taylor S. Chonis\altaffilmark{8,12,17}, C. Martin Gaskell\altaffilmark{8,12}, Evelina R. Gaynullina\altaffilmark{13}, Konstantin N. Grankin\altaffilmark{13,14}, Cecelia H. Hedrick\altaffilmark{8,15}, Mansur A. Ibrahimov\altaffilmark{13}, Elizabeth S. Klimek\altaffilmark{8,16}, Amanda K. Kruse\altaffilmark{8}, Shoji Masatoshi\altaffilmark{8,12}, Thomas R. Miller\altaffilmark{17}, Hong-Jian Pan\altaffilmark{18}, Eric A. Petersen\altaffilmark{8}, Bradley W. Peterson\altaffilmark{8,19}, Zhiqiang Shen\altaffilmark{18}, Dmitriy V. Strel'nikov\altaffilmark{13,20}, Jun Tao\altaffilmark{18}, Aaron E. Watkins\altaffilmark{8}, Kathleen Wheeler\altaffilmark{8}}

\altaffiltext{1}{Institute for Astrophysical Research, Boston University, 725 Commonwealth Avenue, Boston, MA 02215}

\altaffiltext{2}{Astronomical Institute of St. Petersburg State University, Universitetskij Pr. 28, Petrodvorets, 198504 St. Petersburg, Russia}

\altaffiltext{3}{Department of Physics and Astronomy, University of Southampton, Southampton, SO17 1BJ, United Kingdom}

\altaffiltext{4}{Astronomy Department, University of Michigan, 830 Dennison, 501 East University Street, Ann Arbor, Michigan 48109-1042}

\altaffiltext{5}{Mets\"ahovi Radio Observatory, Helsinki University of Technology TKK, Mets\"ahovintie 114,
FIN-02540 Kylm\"al\"a, Finland}

\altaffiltext{6}{Department of Physics and Astronomy, Georgia State University, Atlanta, GA 30303}

\altaffiltext{7}{Department of Physics, Bucknell University, Lewisburg PA 17837}

\altaffiltext{8}{Department of Physics and Astronomy, University of Nebraska, Lincoln, NE 68588-0111}

\altaffiltext{9}{Department of Physics \& Astronomy, University of California, Irvine, CA 92697-4575}

\altaffiltext{10}{Physics Department, Southwestern University, Georgetown, TX 78627-0770}

\altaffiltext{11}{Deceased, Sept. 13, 2008}

\altaffiltext{12}{Department of Astronomy, University of Texas, Austin, TX 78712-0259}

\altaffiltext{13}{Ulugh Beg Astronomical Institute of the Uzbek Academy of Sciences Astronomicheskaya St. 33, Tashkent, 100052, Uzbekistan}

\altaffiltext{14}{Present address: Crimean Astrophysical Observatory, Nauchny, Crimea, 98409 Ukraine}

\altaffiltext{15}{Present address: Astronomy and Astrophysics Department, Pennsylvania State University, 525 Davey Laboratory, University Park, PA 16802}

\altaffiltext{16}{Present address: Astronomy Department, MSC 4500, New Mexico State University, PO BOX 30001, La Cruces, NM 88003-8001}

\altaffiltext{17}{Miller Observatory, 3400 N. 102nd St., Lincoln, NE. 68527}

\altaffiltext{18}{Key Laboratory for Research in Galaxies and Cosmology, Shanghai Astronomical Observatory, Chinese Academy of Sciences, 80 Nandan Rd., Shanghai, 200030 China}

\altaffiltext{19}{Present Address: Department of Physics and Astronomy, Iowa State University, Ames, IA, 50011-3160}

\altaffiltext{20}{Department of Astronomy, National University of Uzbekistan, Tashkent 700095, Uzbekistan}

\begin{abstract}
We present the results of extensive multi-frequency monitoring of the radio galaxy 3C 120 between 2002 and 2007 at X-ray (2-10 keV), optical (R and V band), and radio (14.5 and 37 GHz) wave bands, as well as imaging with the Very Long Baseline Array (VLBA) at 43 GHz. Over the 5 yr of observation, significant dips in the X-ray light curve are followed by ejections of bright superluminal knots in the VLBA images. Consistent with this, the X-ray flux and 37 GHz flux are anti-correlated with X-ray leading the radio variations. Furthermore, the total radiative output of a radio flare is related to the equivalent width of the corresponding X-ray dip. This implies that, in this radio galaxy, the radiative state of accretion disk plus corona system, where the X-rays are produced, has a direct effect on the events in the jet, where the radio emission originates. The X-ray power spectral density of 3C 120 shows a break, with steeper slope at shorter timescale and the break timescale is commensurate with the mass of the central black hole based on observations of Seyfert galaxies and black hole X-ray binaries. These findings provide support for the paradigm that black hole X-ray binaries and both radio-loud and radio-quiet active galactic nuclei are fundamentally similar systems, with characteristic time and size scales linearly proportional to the mass of the central black hole. The X-ray and optical variations are strongly correlated in 3C 120, which implies that the optical emission in this object arises from the same general region as the X-rays, i.e., in the accretion disk-corona system. We numerically model multi-wavelength light curves of 3C 120 from such a system with the optical-UV emission produced in the disk and the X-rays generated by scattering of thermal photons by hot electrons in the corona. From the comparison of the temporal properties of the model light curves to that of the observed variability, we constrain the physical size of the corona and the distances of the emitting regions from the central BH. In addition, we discus physical scenarios for the disk-jet connection that are consistent with our observations.
\end{abstract}


\section{Introduction}
Stellar-mass black hole X-ray binaries (BHXRBs) and active galactic nuclei (AGNs) are both powered by accretion onto a black hole (BH). In many cases these systems emit radiation over several decades of frequency and possess relativistic jets \citep{mir99}. The above similarities in the basic generation of energy and observational properties have led to the paradigm that these two systems are fundamentally similar, with characteristic time and size scales linearly proportional to the mass of the central BH ($\sim$10 M$_\sun$ for BHXRBs and $10^{6}$ to $10^{9}$ M$_\sun$ for AGNs). Although this paradigm has given rise to the expectation that we might test models of AGNs with observations of BHXRBs, such an approach is unjustified until detailed, possibly quantitative connections between BHXRB systems and AGNs become well-established. The comparison of BHXRBs and AGNs is complicated by the fact that a single AGN usually does not show the entire range of properties that we wish to compare. For example, Seyfert galaxies are the AGNs that most resemble BHXRBs, but their radio jets tend to be weak and non-relativistic \citep[e.g.][]{ulv99}. On the other hand, in radio-loud AGNs with strong, highly variable nonthermal radiation (blazars), the Doppler beamed emission from the jet at most wavelengths masks the thermal emission from the accretion disk and its nearby regions.

One well-established property of BHXRBs is the connection between accretion state and events in the jet. In these objects, transitions to high-soft X-ray states are associated with the emergence of very bright features that proceed to propagate down the radio jet \citep{fen04_2,fen09}. Similarly, an accretion disk-jet connection was established in AGNs by \citet{mar02}, who reported a relationship between X-ray and radio events in the radio galaxy 3C~120 (redshift of 0.033). During $\sim$ 3 yr of monitoring of this object, four dips in the X-ray flux, accompanied by spectral hardening, preceded the appearance of bright superluminal knots in the radio jet. This Fanaroff-Riley class I radio galaxy has a prominent radio jet that displays strong variability in flux and jet structure \citep{gom01,wal01}. The jet lies at an angle $\sim 20\degr$ to our line of sight, significantly wider than is typically the case for blazars \citep{jor05}. At optical and X-ray frequencies, 3C~120 possesses properties similar to Seyfert galaxies and BHXRBs, e.g., a prominent iron emission line at a rest energy of 6.4 keV \citep{gra97,zdz01}. Hence, most of the X-rays are produced in the immediate environs of the accretion disk: the corona, a hot wind, or the base of the jet. Since the superluminal knots are disturbances propagating down the jet, a link between decreases in X-ray production and the emergence of new superluminal components demonstrates the existence of a strong disk-jet connection.

In this paper we present the results of extensive multi-frequency monitoring of 3C~120 between 2002 and 2007 at X-ray energies (2.4-10 keV), optical $R$ and $V$ bands, and radio frequencies 37 and 14.5 GHz, as well as imaging with the Very Long Baseline Array (VLBA) at 43 GHz. We use these data to obtain quantitative support for the disk-jet connection claimed by \citet{mar02}. Times of ``ejections" are defined as the extrapolated time of coincidence of a moving knot with the position of the (presumed quasi-stationary) core in the VLBA images. \citet{sav02} have found that ejection times are generally associated with an increase in flux at 37 GHz. Therefore, the 37 GHz flux should be anti-correlated with the X-ray flux if decreases in X-ray production are linked to ejections. Here, we confirm the existence of this anti-correlation and test for a relation between the amplitude of the X-ray dips and that of the associated 37 GHz flares. 

Another similarity between BHXRBs and Seyfert galaxies lies in their X-ray power spectral densities (PSDs). The PSD corresponds to the power in the variability of flux on different timescales. The X-ray PSDs of Seyfert galaxies can be fit by piece-wise power laws with a break \citep{now99,utt02,mch04,mar03,pou01,ede99}. This is generally similar to that of BHXRBs although in many cases the latter have more than one break and sharp peaks due to quasi-periodic oscillations.  We use our data set to verify that the X-ray PSD of 3C~120 also has a break, as inferred by \citet{mar09}, who calculated the PSD from somewhat less extensive observations that partially overlap with ours.

The bulk of the optical continuum from non-blazar AGNs is thought to emanate from the accretion disk \citep{mal83}. In order to test this, we cross-correlate the optical flux variations of 3C 120 with those at X-ray energies. However, the coefficients of raw correlations may be artificially low due to the uneven sampling present in the data. In light of this, we use simulated light curves, based on the underlying PSD, to estimate the significance of the derived coefficients of both the X-ray/optical correlation and X-ray/37 GHz anti-correlation found in the 3C 120 data.

In {\S}2 we present the observations and data reduction procedures, while in {\S}3 we describe the power spectral analysis and its results. In {\S}4 we cross-correlate the X-ray and 37 GHz light curves and quantitatively investigate the relation between the X-ray dips, radio flares and ejections, as well as discuss the X-ray/optical correlation. In {\S}5 we discuss and interpret the results, focusing on the implications regarding the accretion disk-jet connection and the source of the optical emission in 3C~120. {\S}6 presents conclusions, including comparison of 3C 120 with BHXRBs

\section{Observations and Data Analysis}
Table~\ref{data} summarizes the intervals of monitoring at different frequencies for each of the three wave bands in our data set. We term the entire light curve ``monitor data''; shorter segments of more intense monitoring are described below.

The X-ray light curves are based on observations of 3C~120 with the {\it Rossi} X-ray Timing Explorer (RXTE) from 2002 March to 2007 May. We observed 3C~120 with the RXTE PCA instrument with typical exposure times of 1-2 ks. For each exposure, we used routines from the X-ray data analysis software FTOOLS and the program XSPEC to calculate and subtract an X-ray background model from the data and to fit the X-ray spectrum from 2.4 to 10 keV as a power law with low-energy photoelectric absorption by intervening gas in our Galaxy. For the latter, we used a hydrogen column density of $1.23\times 10^{21}$ atoms cm$^{-2}$ \citep{elv89}.

\begin{figure}
\epsscale{0.8}
\plotone{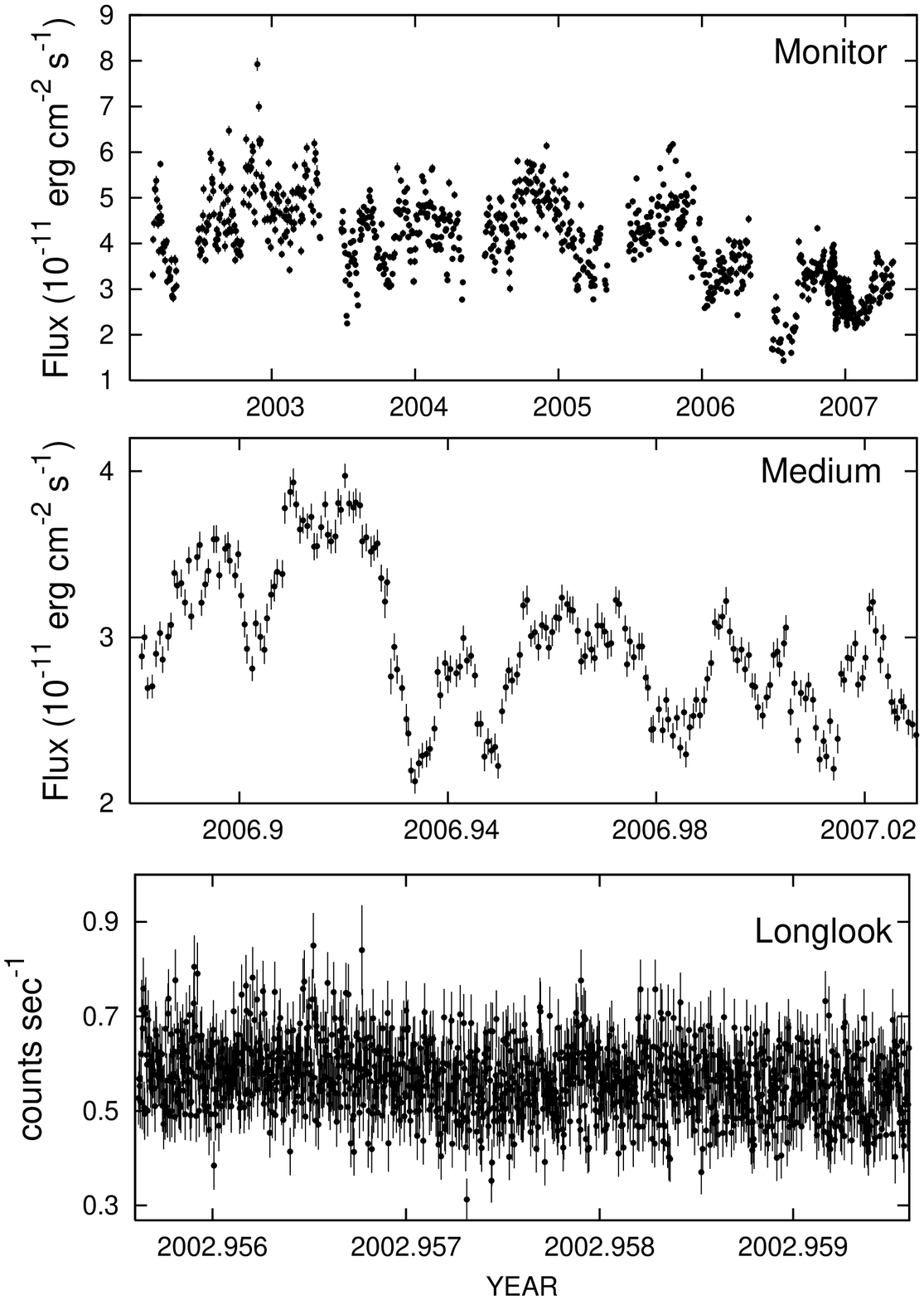}
\caption{X-ray light curves with different sampling rates.}
\label{xdata}
\end{figure}

The sampling of the X-ray flux varied. Normally, observations were made 2-3 times per week except during 8-week intervals each year when the radio galaxy is too close to the Sun's celestial position to observe safely. In order to sample shorter-term variations, between 2006 November and 2007 January we obtained, on average, four measurements per day for almost two months. We refer to these observations as the ``medium'' data. XMM-Newton observed 3C 120 quasi-continuously for about 130 ks on 2002 December 13 and 14, during which all instruments were operating normally. The data were processed with the latest software (SAS version 5.3.3). Light curves were extracted in two energy bands, 0.3-10 keV and 4-10 keV, and were background-subtracted and binned to 100 s time intervals. The 4-10 keV light curve, re-binned to an interval of 0.01 day, were used as the ``longlook" data. We use the 4-10 keV data since the energy range is similar to that of our RXTE data (2.4-10 keV). Figure~\ref{xdata} presents these three data sets. 

We also monitored 3C~120 in the optical $R$ and $V$ bands over a portion of the time span of the X-ray observations. The majority of the measurements in R band are from the 2 m Liverpool Telescope (LT) at La Palma, Canary Islands, Spain, supplemented by observations at the 1.8 m Perkins Telescope of Lowell Observatory, Flagstaff, Arizona. The V-band photometry was obtained with the 0.4-m telescope of the University of Nebraska. On each night a large number of one-minute images ($\sim 20$) were taken and measured separately. Details of the observing and reduction procedure are as described in \citet{kli04}. Comparison star magnitudes were calibrated using \citet{dor06}. To minimize the effects of variations in the image quality, fluxes were measured through an aperture of 8 arcseconds radius. The errors given for each night are the uncertainties in the means. The Miller Observatory observations were taken with a 0.4-m telescope in Nebraska and reduced in similar manner as were the University of Nebraska observations. Observations at the Shanghai Astronomical Observatory were obtained with the 1.56-m telescope at Sheshan Station with standard Johnson-Cousins V, R, and I filters, and all the magnitudes were scaled to the V passband. The reductions were as for the Nebraska observations. Early observations (2002 to 2004) carried out at the Perkins Telescope were in V-band. For these measurements, we used stars D, E, and G from \citet{ang71} to calculate the V-magnitudes. We use three comparison stars in the field of 3C 120 to calculate the R-magnitude. We determined the R-magnitudes of these three stars based on $\sim$20 frames obtained within 2 yr. We use the flux-magnitude calibration of \citet{mea90} and correct for Galactic extinction for both R and V bands.

\begin{figure}
\epsscale{0.8}
\plotone{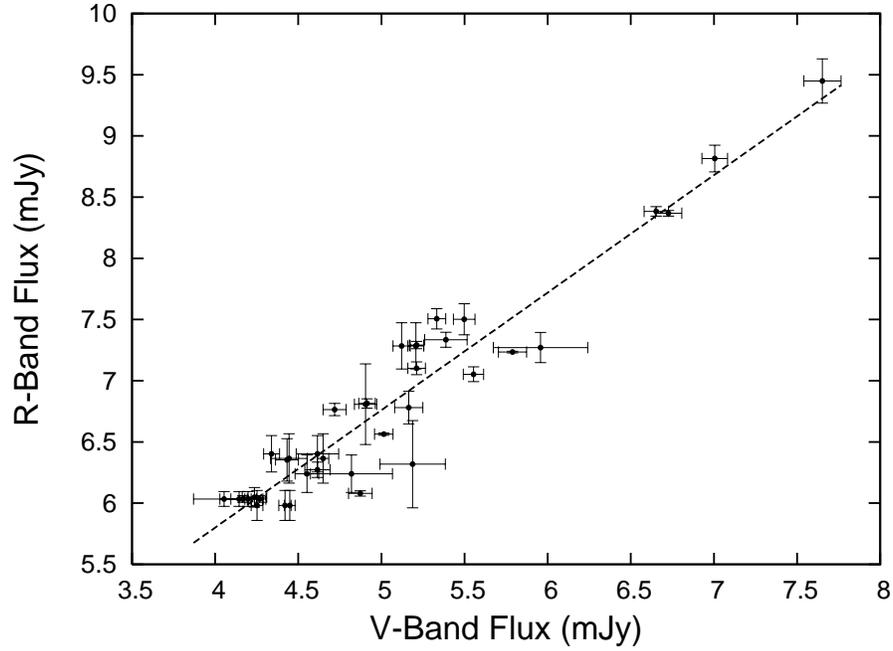}
\caption{Filled circles show the 38 data points that are measured within 0.5 day of each other in V and R band, along with the respective uncertainties. The dashed line represents the best fit straight line through these points, which is used for the V band to R band flux conversion.}
\label{VR_correction}
\end{figure}

\begin{figure}
\epsscale{0.8}
\plotone{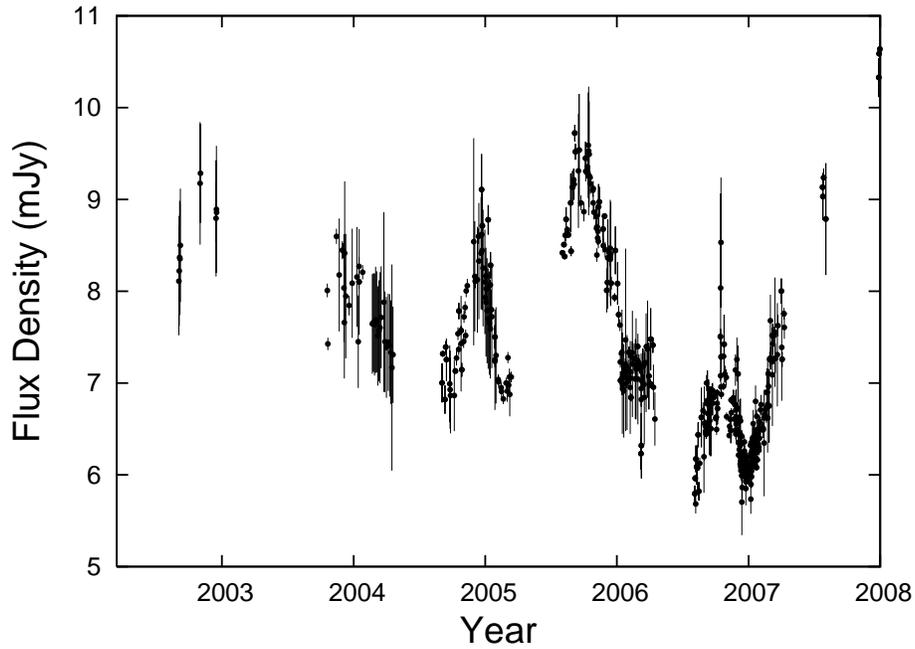}
\caption{Light curve of 3C 120 constructed by combining the V and R band light curves.}
\label{opVR}
\end{figure}

Since the optical sampling in R or V band individually is not as frequent as the X-ray sampling, we construct a better-sampled optical light curve by combining these two bands. We find that the R and V band light curves have 38 data points that were measured within 0.5 day of each other. These data are shown in Figure~\ref{VR_correction}. The equation of the best-fit line is $F_R=0.96F_V+1.96$, where $F_V$ and $F_R$ are the fluxes in V and R band, respectively. We convert the V band fluxes into R band using this equation. We also add some data from Lowell Observatory while constructing the combined light curve, which we present in Figure~\ref{opVR}.

We have compiled a 37 GHz light curve with data from the 13.7 m telescope at Mets\"ahovi Radio Observatory, Finland. The flux density scale is set by observations of DR 21. Sources 3C 84 and 3C 274 are used as secondary calibrators. A detailed description on the data reduction and analysis is given in \citet{ter98}. We also monitored 3C~120 at 14.5 GHz with the 26 m antenna of the University of Michigan Radio Astronomy Observatory. Details of the calibration and analysis techniques are described in \citet{all85}. At both frequencies the flux scale was set by observations of Cassiopeia A \citep{baa77}. 

\begin{figure}
  \begin{center}
    \begin{tabular}{cc}
             \resizebox{50mm}{!}{\includegraphics{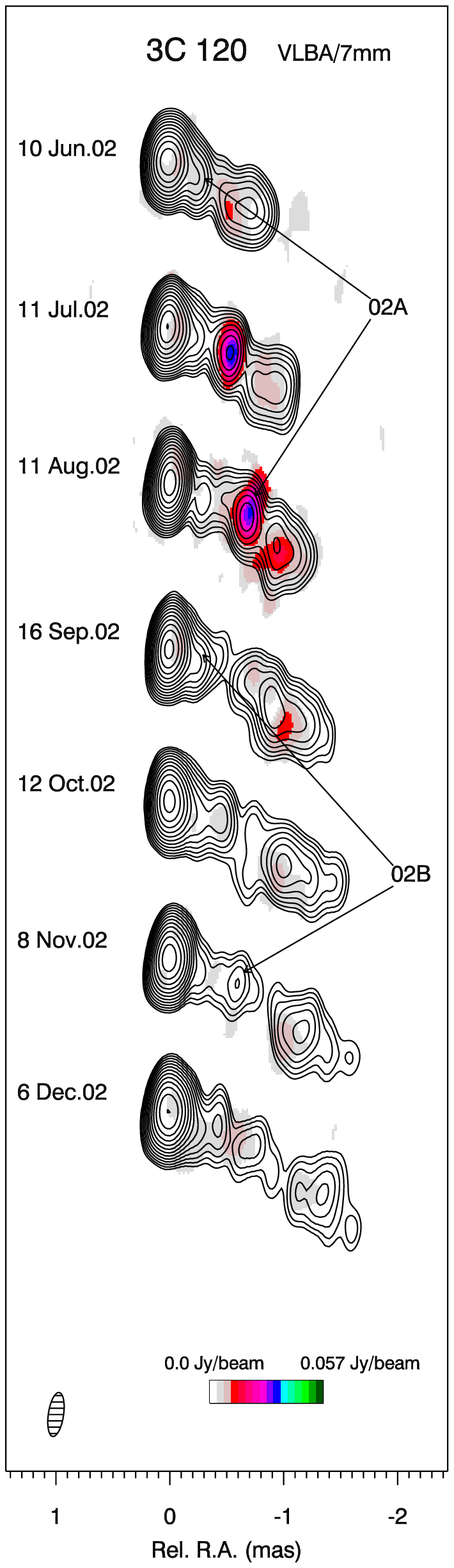}}
             \resizebox{50mm}{!}{\includegraphics{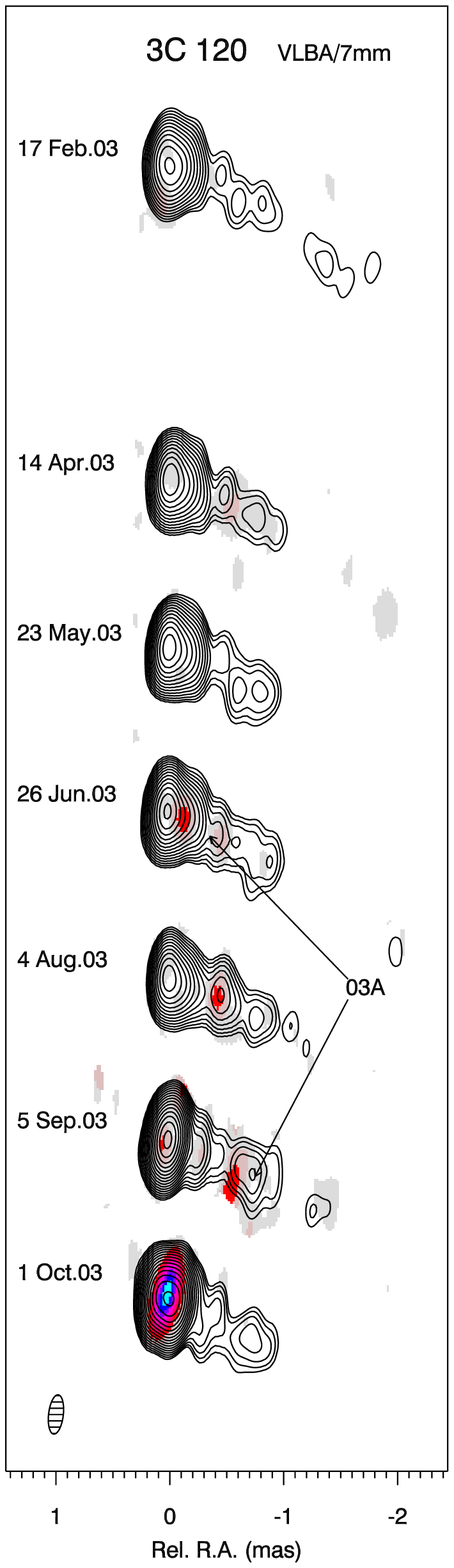}}
             \resizebox{50mm}{!}{\includegraphics{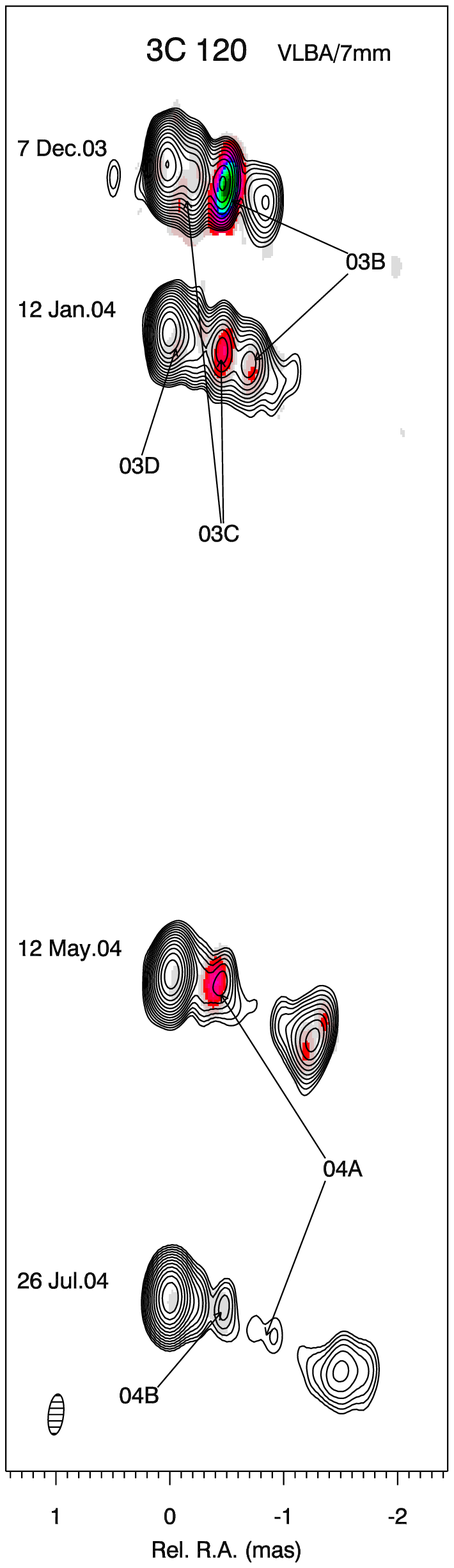}}\\
\end{tabular}
    \caption{VLBI images of 3C 120 at 7 mm obtained from 2002 to 2004. The contours and color show the total and polarized intensity, respectively. The images are convolved with an elliptical Gaussian beam of size 0.36$\times$0.15 mas at PA = $-6\degr$. The global peak over all maps is 1.52 Jy/Beam. The contour levels are 0.25, 0.5, 1.0, ..., 64.0, 90.0\% of the global peak. Individual moving knots are marked at selected epochs.\label{vlba1}}
\end{center}
\end{figure}

\begin{figure}
  \begin{center}
    \begin{tabular}{cc}
             \resizebox{45mm}{!}{\includegraphics{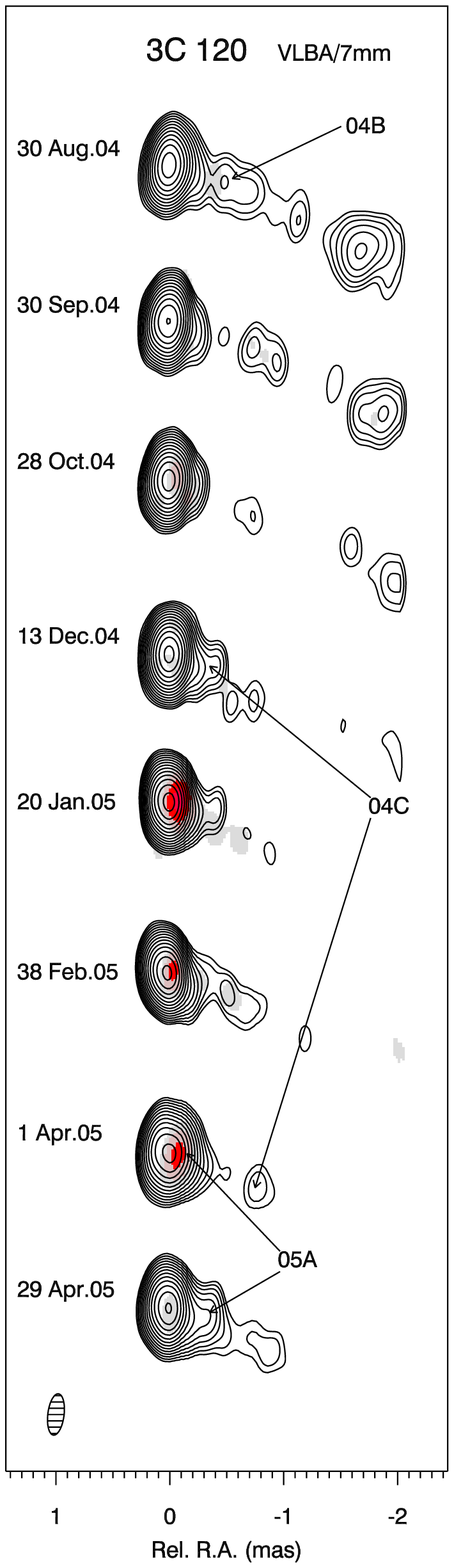}}
             \resizebox{45mm}{!}{\includegraphics{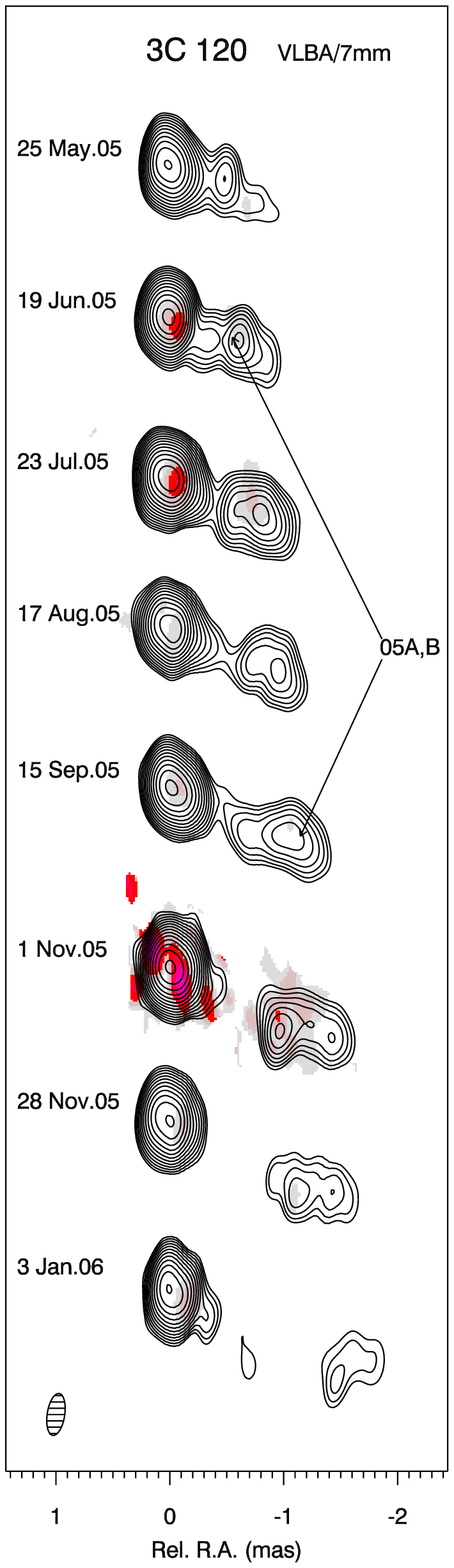}}
             \resizebox{45mm}{!}{\includegraphics{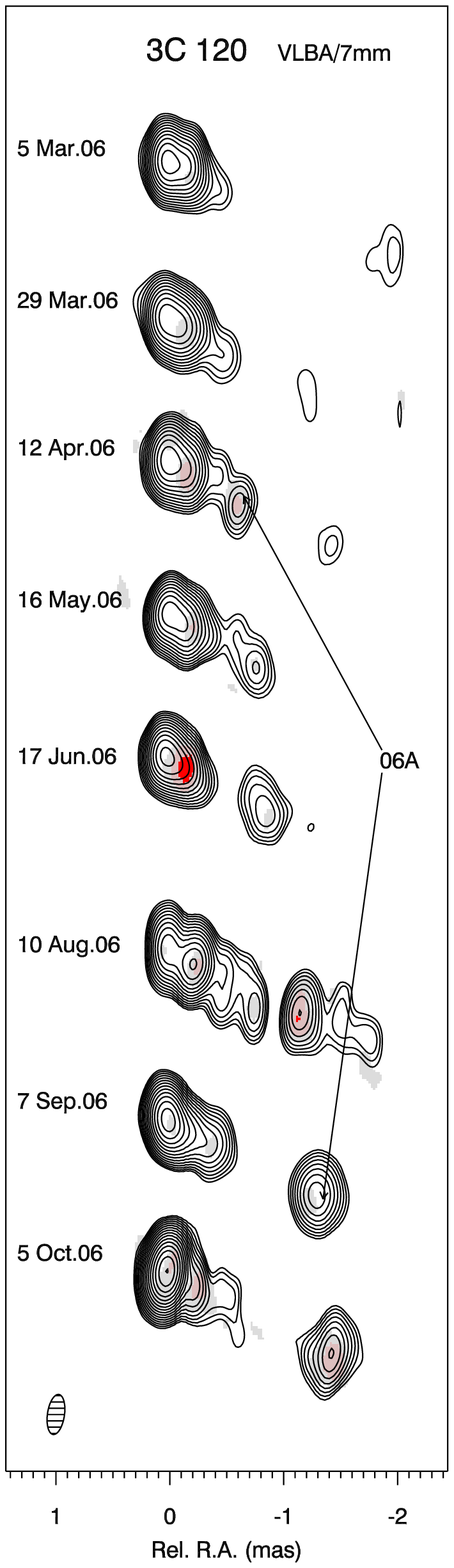}}
             \resizebox{45mm}{!}{\includegraphics{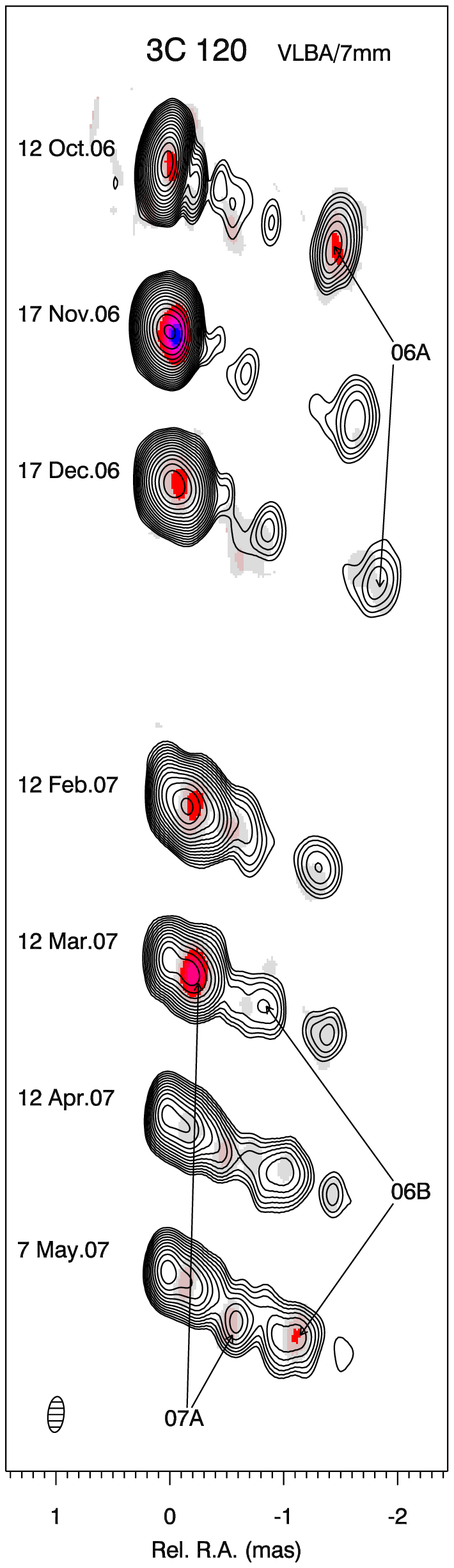}}\\
\end{tabular}
    \caption{VLBI images of 3C 120 at 7 mm obtained from 2004 to 2007. Rest of the description of this figure is the same as Figure \ref{vlba1}.\label{vlba2} }
\end{center}
\end{figure}

\begin{figure}
\epsscale{0.8}
\plotone{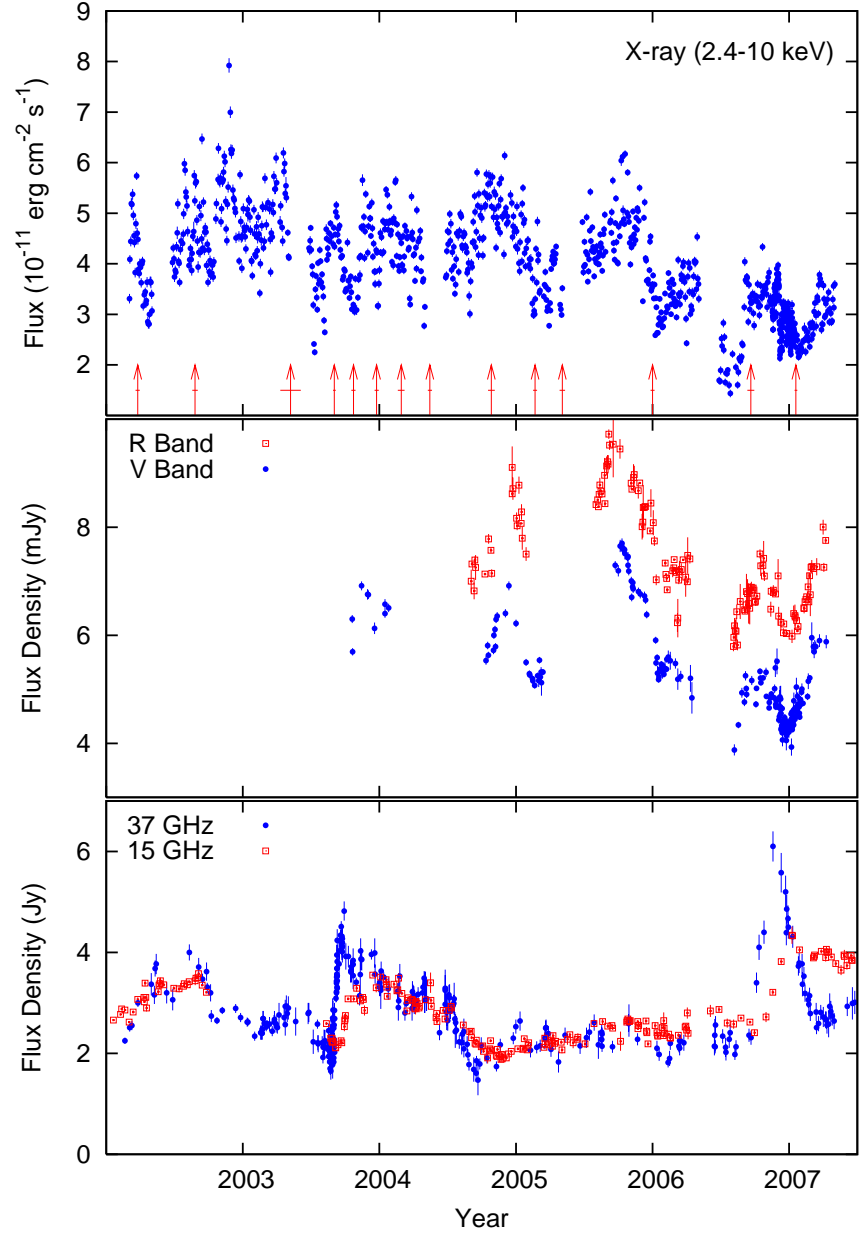}
\caption{Variation of X-ray flux, optical flux density and radio flux density of 3C 120 from 2002 to 2007. In the top panel, the arrows show the times of superluminal ejections and the line segments perpendicular to the arrows indicate the uncertainties in the times.}
\label{xopradspec}
\end{figure}

\begin{figure}
\epsscale{0.8}
\plotone{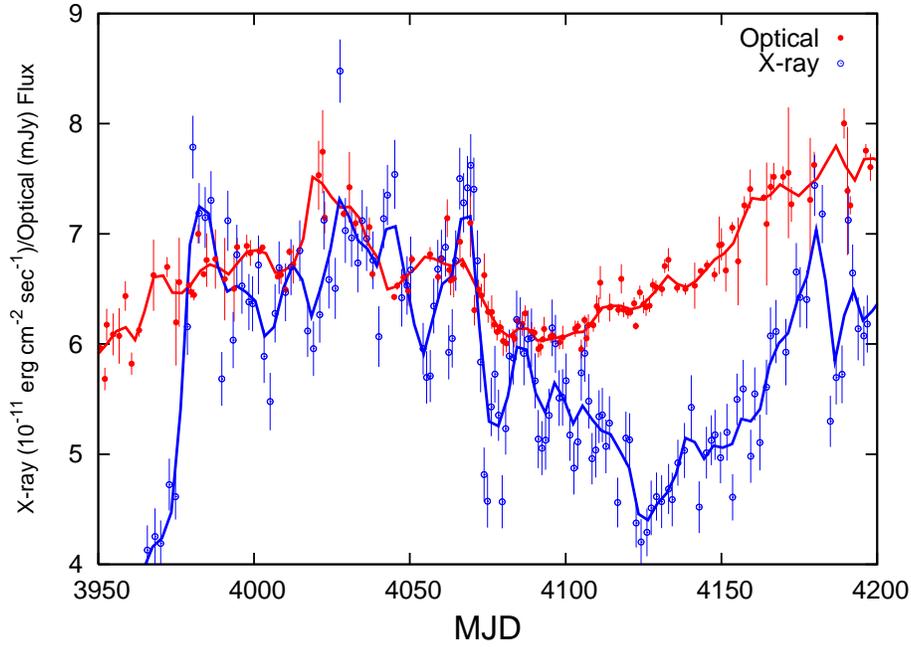}
\caption{Points represent the X-ray and optical light curves between MJD 3950 and 4200 (2006 October and 2007 April) when the time sampling was dense during a minimum in the light curves. The curves represent the same data smoothed with a Gaussian function with a 3 day FWHM smoothing time.}
\label{xop_commondips}
\end{figure}

Starting in 2001 May, we observed 3C~120 with the Very Long Baseline Array (VLBA) at 43 GHz at roughly monthly intervals, with some gaps of 2-4 months. The sequence of images from these data provide a dynamic view of the jet at an angular resolution $\sim$0.1 milliarcseconds (mas) in the direction of the jet, corresponding to 0.064 pc for Hubble constant $H_{0}$ = 70 km s$^{-1}$ Mpc$^{-1}$. We processed the data in the same manner as described in \citet{jor05}. We modeled the brightness distribution at each epoch with multiple circular Gaussian components using the task MODELFIT of the software package DIFMAP \citep{she97}. This represents the jet emission downstream of the core by a sequence of knots (also referred to as ``components''), each characterized by its flux density, FWHM diameter, and position relative to the core. The apparent speeds of the moving components with well-determined motions are all within, $4.0c\pm0.2c$. The ejection time $T_0$ is the extrapolated time of coincidence of a moving knot with the position of the (presumed stationary) core in the VLBA images. Table~\ref{ejecflare} lists the ejection times and Figure \ref{vlba1} and \ref{vlba2} display the VLBA images. Fig.~\ref{xopradspec} presents the X-ray, optical and radio light curves. In the top panel of the figure, the arrows represent the times of superluminal ejections, while the line segments perpendicular to the arrows show the uncertainties in the values of $T_0$. Figure~\ref{xop_commondips} displays the X-ray and optical light curves in the interval between 2006 October and 2007 April. It shows the best-sampled pair of dips in the X-ray and optical fluxes.

\section{Power Spectral Analysis and Results}
We use a variant of the Power Spectrum Response method \citep[PSRESP][]{utt02} to determine the intrinsic PSD of the X-ray light curve. Our realization of PSRESP is described in \citet{cha08}. PSRESP gives both the best-fit PSD model and a ``success fraction'' $F_{\rm succ}$ (fraction of simulated light curves that successfully represent the observed light curve) that indicates the goodness of fit of the model.

\begin{figure}
\epsscale{0.8}
\plotone{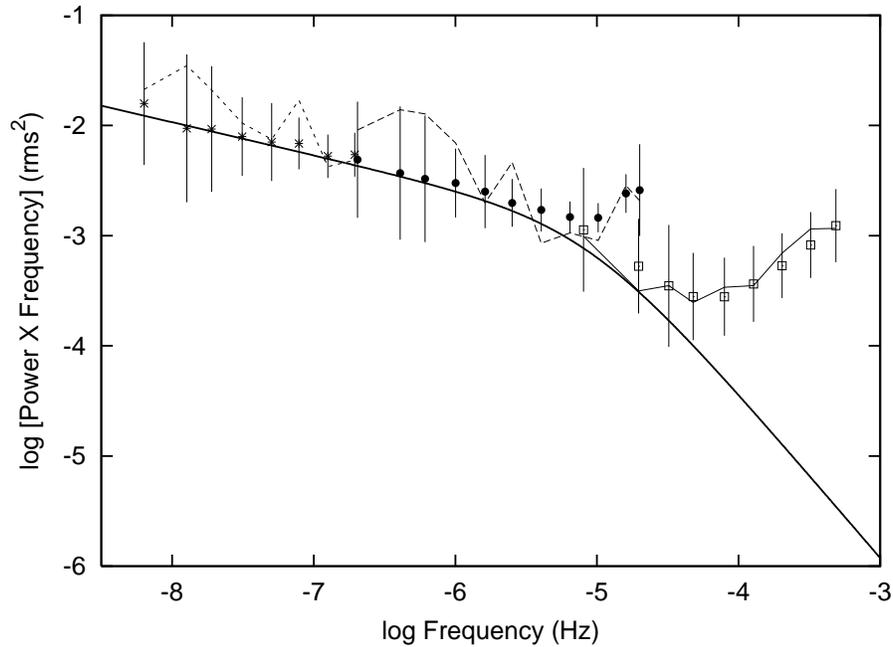}
\caption{Result of application of the PSRESP method to the light curves. The PSD of the observed data at high, medium and low frequencies is given by the solid, dashed and dotted jagged lines, respectively, while the underlying power-law model is given by the thicker solid bent line. Points with error bars (open squares, solid circles and asterisks for high, medium and low frequency range, respectively) correspond to the mean value of the PSD simulated from the underlying power-law model (see text). The errorbars are the standard deviation of the distribution of simulated PSDs. The broadband power spectral density is best described by a bending power law with low frequency slope $-1.3$, high frequency slope $-2.5$ and break frequency $10^{-5.05}$ Hz.}
\label{psd}
\end{figure}

At first we fit a simple power-law model to the X-ray PSD, but found that the value of $F_{\rm succ}$ was unacceptably low. This implies that a simple power-law is not the best model for this PSD. Then we fit a bending power-law model (broken power law with a smooth break) to the X-ray PSD,
\begin{equation}
P(\nu)=A\nu^{\alpha_L}[1+(\frac{\nu}{\nu_{B}})^{(\alpha_{L}-\alpha_{H})}]^{-1}.
\end{equation}
Here, $A$ is a normalization constant, $\nu_{B}$ is the break frequency, and $\alpha_H$ and $\alpha_L$ are the slopes of the power-laws above and below the break frequency, respectively \citep{mch04}. During the fitting, we varied $\nu_{\rm B}$ from $10^{-9}$ to $10^{-5}$ Hz in steps of $10^{0.05}$, $\alpha_{\rm H}$ from $-1.5$ to $-3.0$ in steps of 0.1, and $\alpha_{\rm L}$ from $-1.0$ to $-1.5$ in steps of 0.1. These ranges include the values of $\alpha$ found in the light curves of BHXRBs in the high-soft state, for which $\alpha_{\rm L} \approx -1$ and  $\alpha_{\rm H}$ between $-2$ and $-3$ \citep[e.g.][]{rem06}. This procedure yields a much higher success fraction than the simple power-law model. Based on the model with the highest success fraction, we obtain a best fit with the parameters $\alpha_{\rm L}=-1.3^{+0.2}_{-0.1}$, $\alpha_{\rm H}=-2.5^{+0.3}_{-0.5}$, and log$_{10}(\nu_{\rm B})=-5.05^{+0.2}_{-0.6}$ Hz. The success fraction for this fit is high, $0.9$. The intrinsic scatter of power is very high for slopes below $-2$ due to red noise leak. As a result, trial PSD models with a high frequency slope of any value between $-2$ and $-3$ give very similar success fractions. This causes the larger uncertainty in the value of the high frequency slope relative to that at lower frequencies. Figure~\ref{psd} presents this best-fit model and the corresponding PSD. As seen in the figure, the high frequency part of the PSD is dominated by Poisson noise. That is because this part of the PSD is generated from the longlook light curve, and fluxes in the longlook light curve have larger uncertainties owing to shorter exposure times than those in the other light curves. The figure shows that when the estimated Poisson noise is added to the best-fit model PSD, it matches the observed PSD quite well. The nature of the X-ray PSDs and the values of break frequencies found previously in Seyfert galaxies are consistent with those of BHXRBs in high-soft state \citep{mch05}. This is consistent with accretion rates in all these objects which are above 10\%, more than the typical value of 2\% indicating the transition from low to high state in galactic BHXRBs \citep{mac03}. In 3C 120, the accretion rate $\sim$30\% which implies that it is also a high state system.

The PSD break frequency in BHXRBs and Seyfert galaxies scales inversely with the mass of the black hole while being proportional to the accretion rate \citep{utt02,mch04,mch06,ede99,mar03}. Using the best-fit values and uncertainties in the relation between break timescale, BH mass, and accretion rate obtained by \citet{mch06}, we estimate the expected value of the break frequency in the X-ray PSD of 3C~120 to be $10^{-5.0\pm0.7}$ Hz. Hence, our derived break frequency lies within the expected range. Here we adopt a BH mass of $5.5 \times 10^7$ M$_\sun$ from emission-line reverberation mapping \citep{pet04} and a bolometric luminosity of the big blue bump of $2.2\times 10^{45}$ ergs s$^{-1}$ \citep{woo02}. 

We note that the high and low frequency slopes, as well as the break frequency calculated by \citet{mar09} ($-2.1\pm0.4$, $-1.2$, $10^{-5.75\pm0.43}$ Hz), are consistent with the best-fit values in this paper within the uncertainties. The difference in the central values probably results from our use of a different longlook light curve, which is from the EPIC instrument onboard XMM-Newton. The sensitivity of this detector is higher than the PCA of RXTE, allowing our light curve to have a much better sampling rate. This is useful for extending the lower end of the frequency range of the PSD, leading to a more accurate measurement of the high frequency slope and break frequency.

\section{Cross-correlation Analysis and Results}
We employ the discrete cross-correlation function \citep[DCCF][]{ede88} method to search for connections between variations at pairs of wave bands. We determine the significance of the correlations by performing the analysis on $100$ simulated light curves generated randomly with a PSD matching that of the real data. Details of this procedure are described in \citet{cha08}. In the present case, we correlate simulated X-ray light curves with the observed optical/radio light curves to calculate the significance.

\subsection{X-ray/Radio Correlation}
\begin{figure}
\epsscale{0.8}
\plotone{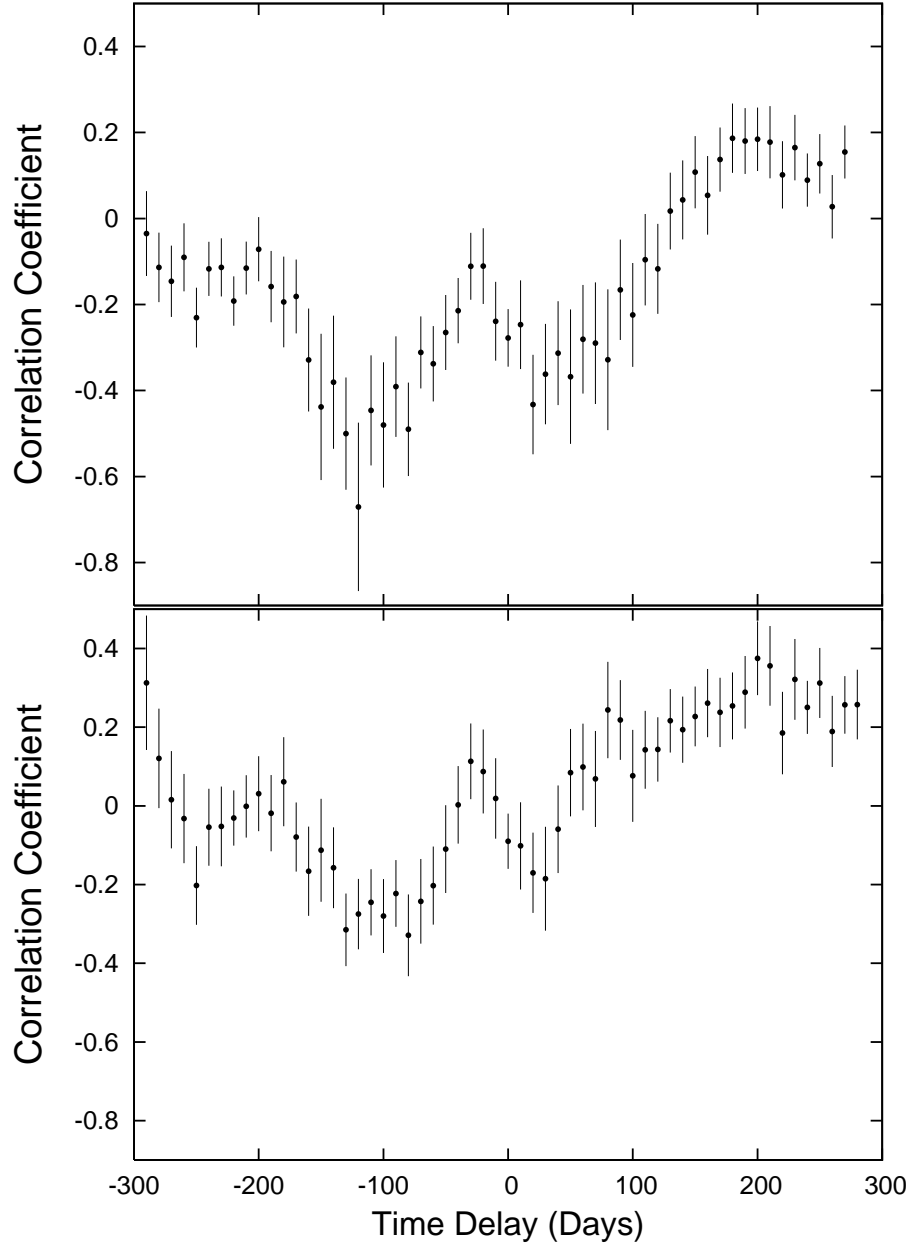}
\caption{Discrete cross-correlation function (DCCF) of the X-ray and radio monitor data. The time delay is defined as positive if the X-ray variations lag those at radio frequency. Top panel shows the correlation function for the entire data set and the bottom panel shows the same excluding the data during the major flare at 37 GHz in 2006-07 and the corresponding deep dip at X-ray energies.}
\label{xradcor}
\end{figure}

The 37 GHz light curve has an average sampling frequency of about once per week. We bin the X-ray and the 37 GHz light curves in 7-day intervals before performing the cross-correlation so that the light curves being compared have similar sampling frequency. As determined by the DCCF (top panel of Figure~\ref{xradcor}), the X-ray flux variations are anti-correlated with those at 37 GHz in 3C~120. The highest amplitude of the  X-ray versus 37 GHz DCCF is at a value of $-0.68\pm0.11$, which corresponds to a 90\% significance level. The time lag of the peak indicates that the X-ray lead the radio variations by $120\pm30$ days. We have used the FR-RSS technique proposed by \citet{pet98} to calculate the mean value and uncertainty of the cross-correlation time lag. This procedure gives quantitative support to the trend that is apparent by inspection of the light curve, i.e., X-ray dips are followed by appearance of new superluminal knots and hence enhancement in the 37 GHz flux. 

The bottom panel of Figure~\ref{xradcor} shows the X-ray/37 GHz DCCF but without the data after 2006 April, in order to exclude the two deepest X-ray dips and highest amplitude flare at 37 GHz toward the end of our monitoring program. The X-ray/37 GHz anti-correlation remains, although the minimum of the DCCF is smaller, $-0.4$, with a significance level of 72\%. The X-ray variations lead the radio by $80\pm30$ days. Thus, the X-ray/37 GHz anti-correlation is not just the consequence of a singular event. The longer time delay of the major event in late 2006 was caused by the longer time between the start of the outburst and the peak relative to other radio flares, especially the one at 2003.7. This is naturally explained as a consequence of higher optical depth of stronger radio outbursts.

\subsection{X-ray Dips and Radio Flares}
In order to determine the physical link between the accretion disk and jet, we check whether the amplitudes of the X-ray dips and associated radio flares are related. To test this, we calculate the equivalent width of each X-ray dip and the area under the curve of each 37 GHz flare to measure the total energy involved in the events. We approximate that the radio light curve is a superposition of a constant baseline of 1.5 Jy from the more extended jet and long-term flares. The baseline is chosen as approximately the lowest level of the 37 GHz light curve. We then follow \citet{val99} by decomposing the baseline-subtracted light curve into individual flares, each with exponential rise and decay. We use four parameters to describe each flare: the rise and decay timescales, and the height and epoch of the peak. Before the decomposition, we smooth the light curve using a Gaussian function with a 10-day FWHM smoothing time. The details of the decomposition procedure are described in \citet{cha08}.

The X-ray light curve has a long-term trend, i.e., the baseline is not constant. We define the baseline X-ray flux as a cubic-spline fit of the annual mean plus one standard deviation. Although this is an arbitrary definition, this baseline reproduces reasonably well the mean flux level in between obvious dips and flares. There is a long-term trend in the spectral index variations as well. The X-ray spectral index $\alpha_x$, defined by $f_x \propto \nu^{\alpha_x}$, where $f_x$ is the X-ray flux density and $\nu$ is the frequency, varied between $-0.5$ and $-1.1$, with an average value of $-0.83$ and a standard deviation of 0.10 over the 5.2 yr of observation. In addition to the short term fluctuations, there is a long-term trend of increasing values of $\alpha_x$ during this interval. We calculate a baseline in the same manner as for the X-ray flux variations to highlight the change in spectral index during the dips (see Figure~\ref{baseline1sig}). Before we perform the above analysis to calculate the baseline, we smooth the X-ray light curve using a Gaussian function with a 10-day FWHM smoothing time. 

\begin{figure}
\epsscale{0.8}
\plotone{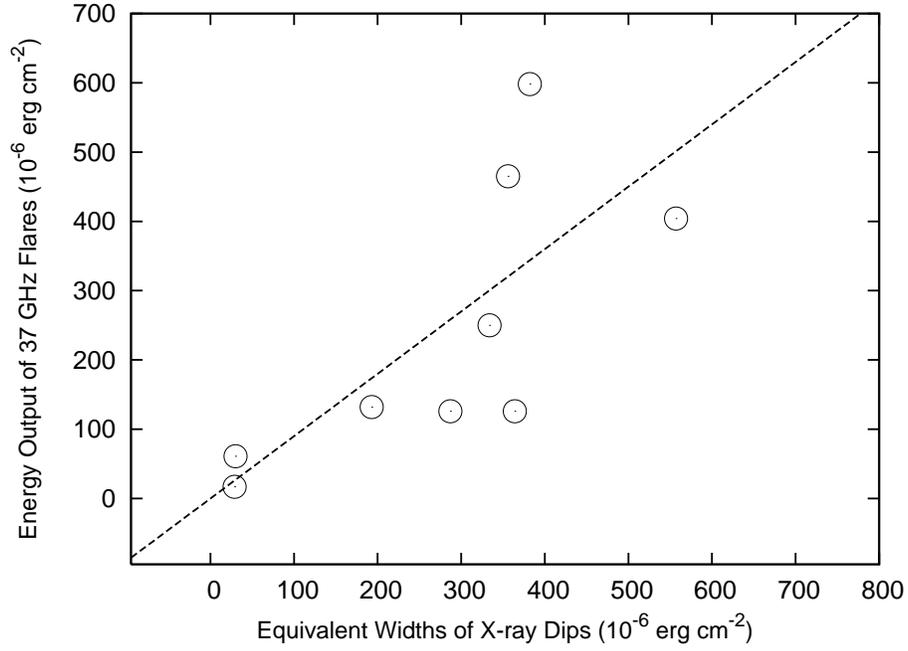}
\caption{Total energy output of the 37 GHz flares versus equivalent width of the corresponding X-ray dips.}
\label{dipejecarea}
\end{figure}

We proceed by subtracting the baseline and then modeling the dips as inverted exponential flares. Figure~\ref{xradmodelfit_1sig} shows the model fits to both the X-ray and 37 GHz light curves. Table~\ref{ejecflare} lists the parameters of the X-ray dips and the corresponding 37 GHz flares along with the times of superluminal ejections. Since the correlation functions with and without the data in 2007 (during the large amplitude 37 GHz flare) show that the X-ray dips lead the 37 GHz flares by $120\pm30$ and $80\pm30$ days, respectively, and we have identified 15 significant dips during the 5.2 yr of monitoring, we assume that a radio flare that peaks between 20 and 180 days after an X-ray minimum is physically related to it. Using the results of this analysis, we plot the equivalent width of X-ray dips versus the energy output of the corresponding 37 GHz flares in Figure~\ref{dipejecarea}. The 37 GHz flare at 2003.72 has a wide decaying wing and there are two X-ray dips (at 2003.58 and 2003.82) that may be related to this flare as well as another smaller 37 GHz flare (at 2003.92) on top of that wing. Similarly, there are two X-ray dips at 2005.12 and 2005.39, and there are four corresponding 37 GHz flares very close to each other in time at 2005.23, 2005.36, 2005.57, 2005.80. In the above cases, we plotted the total energy output of the group of 37 GHz flares against the total equivalent width of the corresponding group of X-ray dips. The energy output of the flares and the dips were corrected by adding the residuals shown in Figure~\ref{xradmodelfit_1sig}. The plot in Figure~\ref{dipejecarea} reveals a positive correlation between the flare and dip strengths. The level of change in the accretion disk and/or corona is therefore closely related to the amount of excess energy injected into the jet. 

\begin{figure}
\epsscale{0.8}
\plotone{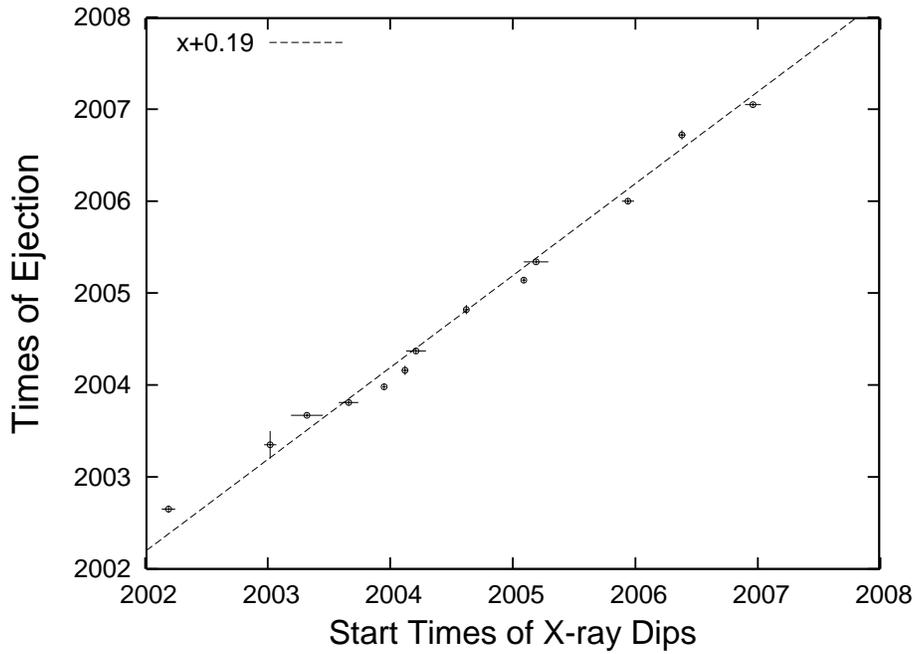}
\caption{Times of ejection of new VLBA knots versus times of start of X-ray dips from Table \ref{ejecflare}. The dashed line is the best-fit straight line through the points. The y-intercept of this line indicates the value of mean time delay between the start of dips and times of ejections.}
\label{dip_ejec_times}
\end{figure}

It can be seen from Table~\ref{ejecflare} that 14 out of 15 significant X-ray dips are followed by a superluminal ejection. The time delay between the start of the X-ray dips and the time of ``ejection" of the corresponding superluminal knot varies between $0.03$ yr to $0.46$ yr, with a mean value of $0.18\pm0.14$ yr. We plot the times of ejection of new knots along with the corresponding times of start of the X-ray dips in Figure \ref{dip_ejec_times}. The uncertainties in the start time are proportional to the widths of the dips and those in the ejection times are as given in Table~\ref{ejecflare}. A straight line fits the data extremely well with small scatter, which indicates that there is a clear association between X-ray dips and superluminal ejections. The best-fit line through the points has a slope of 1 and y-intercept of $0.19$, which is consistent with the mean delay of $0.18\pm0.14$ yr given above. This strongly supports the proposition \citep{mar02} that in 3C 120, a decrease in the X-ray production is linked with increased speed in the jet flow, causing a shock front to subsequently move downstream. Generally, there is a close correspondence of superluminal ejections with 37 GHz flares, but sometimes the decrease in flux farther out in the jet offsets much of the flux increase from the appearance of a new knot. In such cases, the corresponding increase of flux was not large enough to be detected in our decomposition of the smoothed light curve. This causes the minimum in the X-ray DCCF to be less significant than it would if old knots were to completely fade before new ones appear. Despite this complication, the X-ray/37 GHz cross-correlation ({\S}4.1) verifies with an objective statistical method that radio events in the jet are indeed associated with X-ray dips. The apparent speeds of the moving components with well-determined motions are all similar, $4.0c\pm0.2c$. Therefore, a knot moves a distance of $0.22$ pc in $0.18$ yr, projected on the plane of the sky. Since the angle of the jet axis of 3C 120 to the line of sight $\sim$20$^{\circ}$, the actual distance traveled by the knot $\sim0.5$ pc. Hence, we derive a distance $\sim0.5$ pc from the corona (where the X-rays are produced) to the VLBA 43 GHz core region. This confirms that the core is offset from the position of the BH. This is one of the few cases where we are able to probe the upstream region of the core using the time variable emission at a combination of wave bands.

\subsection{X-ray/Optical Correlation}
\begin{figure}
\epsscale{0.8}
\plotone{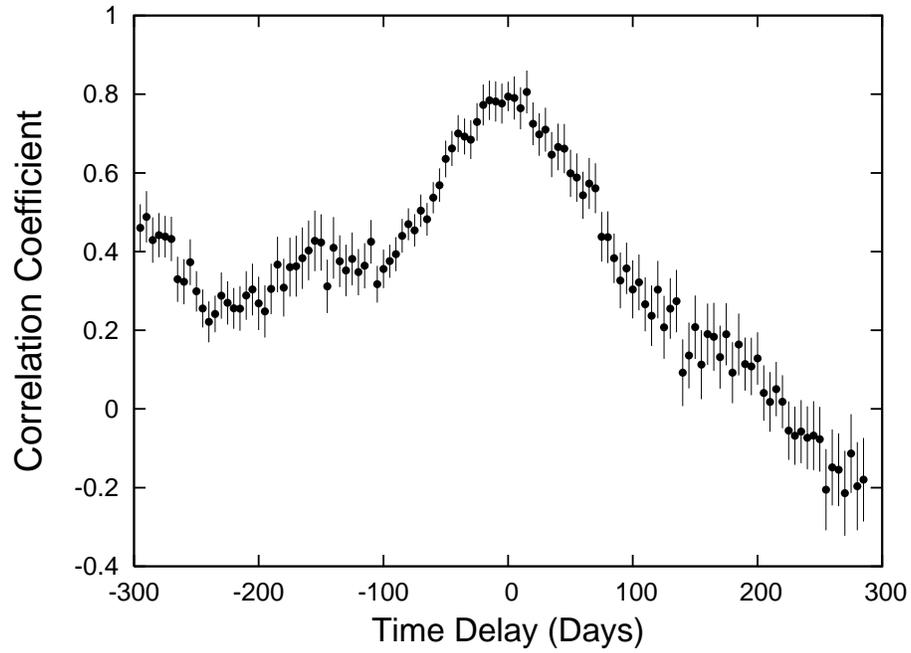}
\caption{Discrete cross-correlation function (DCCF) of the optical and X-ray monitor data for the entire 5 yr interval. The time delay is defined as positive if the variations at the higher frequency waveband lag those at the lower frequency.}
\label{xop}
\end{figure}

We bin the X-ray and the combined optical light curves in 2-day intervals before performing the cross-correlation, so that the light curves being compared have similar sampling frequency. As determined by the DCCF (Figure~\ref{xop}), we find that the X-ray variations are very strongly correlated with those at optical wavelengths in 3C 120. The peak X-ray versus optical DCCF is $0.80\pm0.07$, which corresponds to a 99\% significance level. The position of the peak of the correlation function indicates the relative time delay between the variations at the two wavelengths. In this case, the peak is very wide, so that the value of the relative time delay can not be easily estimated from the DCCF plot. We used the FR-RSS technique proposed by \citet{pet98} to calculate the mean value and the uncertainty of the cross-correlation time lag. This method indicates that the X-ray variations lead the optical by $0.5\pm4$ days. The highly significant correlation and short time delay between the X-ray and optical variations indicates that emission at these wave bands is at least partially co-spatial. The asymmetry in the correlation function at time delays above $\sim$100 days indicate that the X-ray variations lead those in the optical at these longer timescales.

To characterize the variation of the X-ray/optical time lag over the years, we divide both light curves into two intervals, 2004 July to 2005 May and 2005 June to 2007 May, and repeat the DCCF analysis on each segment. The result (Figure~\ref{xop_tw}) indicates that in the first segment, the X-ray variations lead those in the optical by $\sim$25 days while in the second segment the correlation function is similar to what we obtained for the entire time interval, with similar non-zero time delay (Figure~\ref{xop}). This variation of the time lag over the years may be the cause of the observed wide peak in the correlation function.

\begin{figure}
\epsscale{0.8}
\plotone{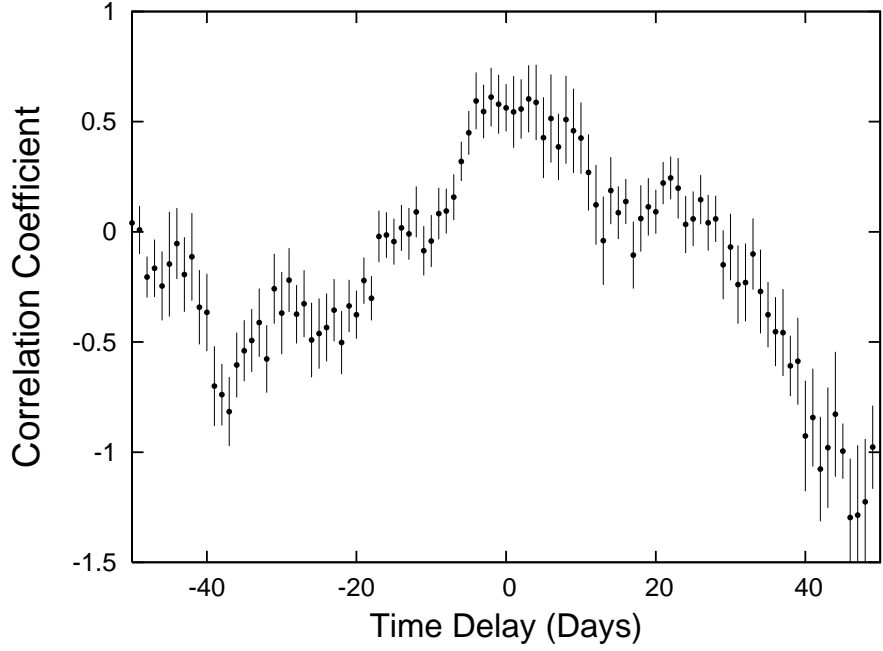}
\caption{Discrete cross-correlation function (DCCF) of the optical and X-ray data between 2006 November and 2007 January, binned to an interval of 0.5 day.}
\label{xop_small}
\end{figure}

Between 2006 November and 2007 January, the X-ray light curve (``medium" light curve described in {\S}2) was sampled 4 times per day and the combined optical light curve has a sampling rate of twice per day, on average. We bin these light curves in 0.5 day intervals. We cross-correlate the binned light curves in order to compare the correlation function with that of the long-term light curves. The correlation function, shown in Figure~\ref{xop_small}, has a similarly significant correlation coefficient and time delay. The similar values of correlation coefficient and time delay using very well-sampled light curves illustrates the robustness of the correlation result.

The X-ray to optical time delay calculated by \citet{mar09} has the same sign as determined in this paper, although the magnitude is significantly different ($28.73^{+6.19}_{-5.87}$ days). We note that (1) This work includes more data points in the R band light curve, including the period $2007.0-2007.36$, and 2) We construct a combined light curve using the R band data as well as scaled V band data. We use this more completely sampled light curve for the correlation analysis. This, in addition to the longer time lag from 2004.5 to 2005.5, can account for the difference between our result and that obtained by \citet{mar09}.

The interpolated cross-correlation function (ICCF) is another method to calculate the cross-correlation of unevenly sampled discrete data \citep{gas87}. We have repeated all the above correlation analysis using the ICCF and have found that the DCCF and ICCF results are consistent with each other for the data presented in this work.

\begin{figure}
\epsscale{0.8}
\plotone{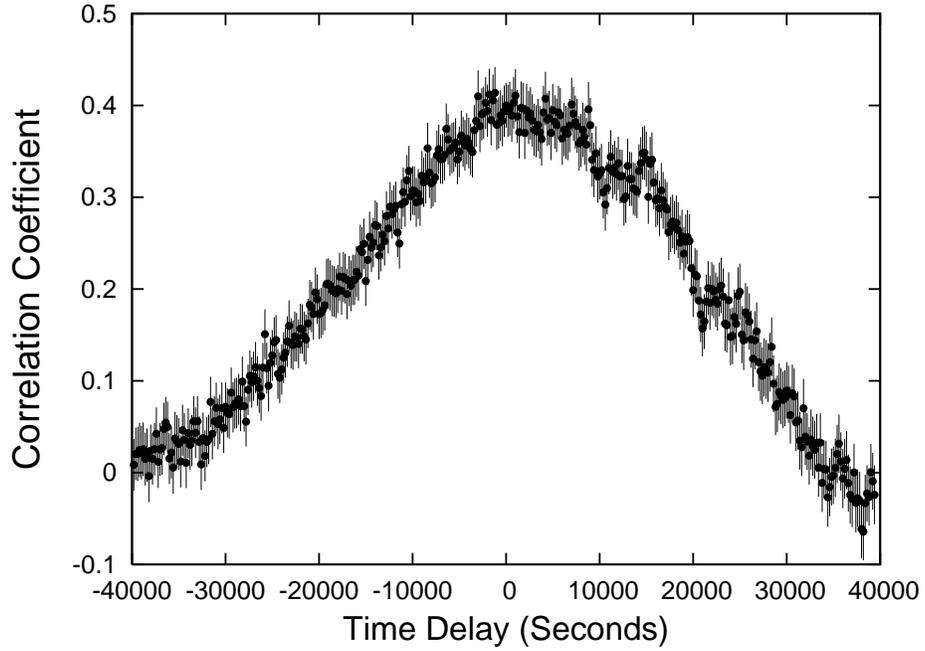}
\caption{Discrete cross-correlation function (DCCF) of the soft and hard longlook X-ray data. The time delay is defined as positive if the variations at the hard wave band lag those at the soft wave band.}
\label{softhard}
\end{figure}

We have cross-correlated the hard (4-10 keV) and soft (0.2-4 keV) longlook X-ray light curves, finding an excellent correlation with nearly zero time lag (Figure~\ref{softhard}). This indicates that in 3C 120, variations in soft and hard X-rays follow each other very closely.

\section{Discussion}

\subsection{Disk-Jet Connection}
The physical cause of the connection between events in the central engine and the jet of BHXRBs and AGNs is currently a matter of considerable speculation. If the jet is magnetically launched from the accretion disk \citep{bla82}, then there must be a link between the magnetic state at the base of the jet and the accretion state in the inner disk. One scenario, proposed for BHXRBs by \citet{liv03} and \citet{kin04}, involves a change in the magnetic field configuration in the inner disk from a turbulent condition in the high-soft state (when the X-ray flux is relatively higher and softer) to mainly poloidal in the low-hard X-ray state. The turbulence is needed for viscous heating, which in the BHXRB case leads directly to bright X-ray emission with a soft spectrum. In an AGN it causes strong ultraviolet emission, which is Compton scattered in the corona to a hard X-ray spectrum. If the field switches from chaotic to mainly poloidal, which \citet{liv03} suggest can occur by random episodes of near-alignment of the field in the relatively small number of turbulent cells in the inner disk, then the radiation in the inner disk will be quenched at the same time as energy flow into the jet is promoted. The transition back to the turbulent, radiative inner disk of the high-soft state would need to involve a surge of energy injected into the jet in order to send a shock wave \citep{mil05} down the jet. Perhaps global magnetic reconnection could cause this, but no detailed MHD model has been published to date. 

Alternatively, it is possible that the ``corona," where the X-ray emission seen in AGN supposedly arises from Compton up-scattering of softer accretion-disk photons, might be the base of the jet \citep{mar05}. If this is the case, then the X-ray flux will be related to the number of electrons residing there and available for scattering to create X-rays. The mass loading of the jet should also affect the asymptotic Lorentz factor of the flow downstream if the jet is magnetically driven \citep[e.g.][]{vla04}. The same decrease in electron number that causes a drop in scattered X-ray emission near the disk would lead to a time-delayed increase in the speed of the jet downstream. The flatter-spectrum nonthermal X-ray emission from the downstream jet would then play a larger relative role in the X-ray emission, causing the observed hardening of the spectrum during the dips. The increase in flow speed of the jet could form a shock wave, seen as a superluminal radio knot. It is difficult to speculate why the mass loading should change, since we do not understand the processes by which material from the disk and/or ergosphere are injected into the base of the jet. However, observations of the microquasar GRS 1915+105 suggest that outflow of matter from the disk switches from mainly a wind to the jets \citep{nei09}. In an AGN with a magnetically driven jet, it may be the case that lower mass injection into the jet actually enhances the jet emission owing to an increase in the flow speed driving a shock wave down the jet, as described above. 

\subsection{Source of Optical Emission}
The strong correlation between the X-ray and optical variations in 3C 120 implies that the emission at both wave bands arises from the same general region. Since the X-rays originate in the corona, the optical emission is probably thermal emission from the accretion disk \citep{mal83}. We can reject the alternative hypothesis that the main component of optical emission is synchrotron radiation from the jet. In that case, the emission should be significantly polarized, contrary to observations showing the optical linear polarization to be $<0.3\%$ \citep{jor07}. Furthermore, the anti-correlation of the X-ray and 37 GHz emission (the latter of which is produced in the jet) implies that any optical synchrotron emission from the jet should also anti-correlate with the X-ray flux, contrary to the strong observed correlation.

Although the X-rays are predominantly produced by inverse Compton (IC) scattering of the thermal optical/UV seed photons from the accretion disk by hot, but non-relativistic, electrons in the corona, the optical/UV emission and the X-rays are tied together by another mechanism: some of the optical/UV radiation is produced by heating of the accretion disk by X-rays produced in the above process (``feedback" mechanism). The amplitude of the short timescale ($\sim$days) variations in the X-ray light curve is larger than that in the optical. This can happen if a significant fraction of the optical emission is due to heating of the disk by X-rays, since this reprocessing may smooth out the short timescale variations. Feedback can make the time delay more difficult to define by producing a fraction of the optical/UV photons with different temporal properties from that of the direct emission.

\begin{figure}
\epsscale{0.8}
\plotone{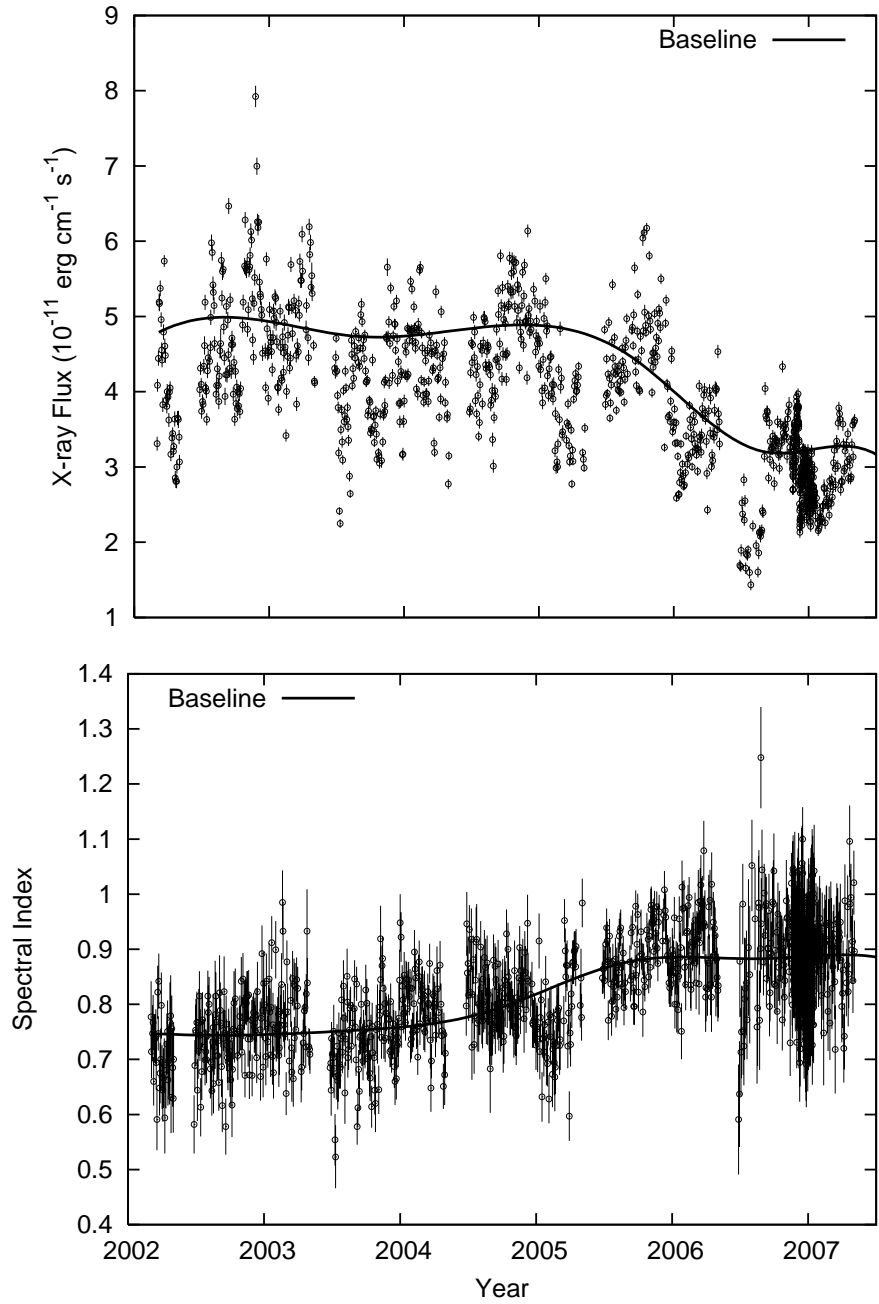}
\caption{Variation of X-ray flux and the X-ray spectral index during 2002-2007. Curves are the spline interpolation of the yearly mean of the data plus one standard deviation. In case of X-rays, variations (both positive and negative) are relative to the baseline.}
\label{baseline1sig}
\end{figure}

\begin{figure}
\epsscale{0.8}
\plotone{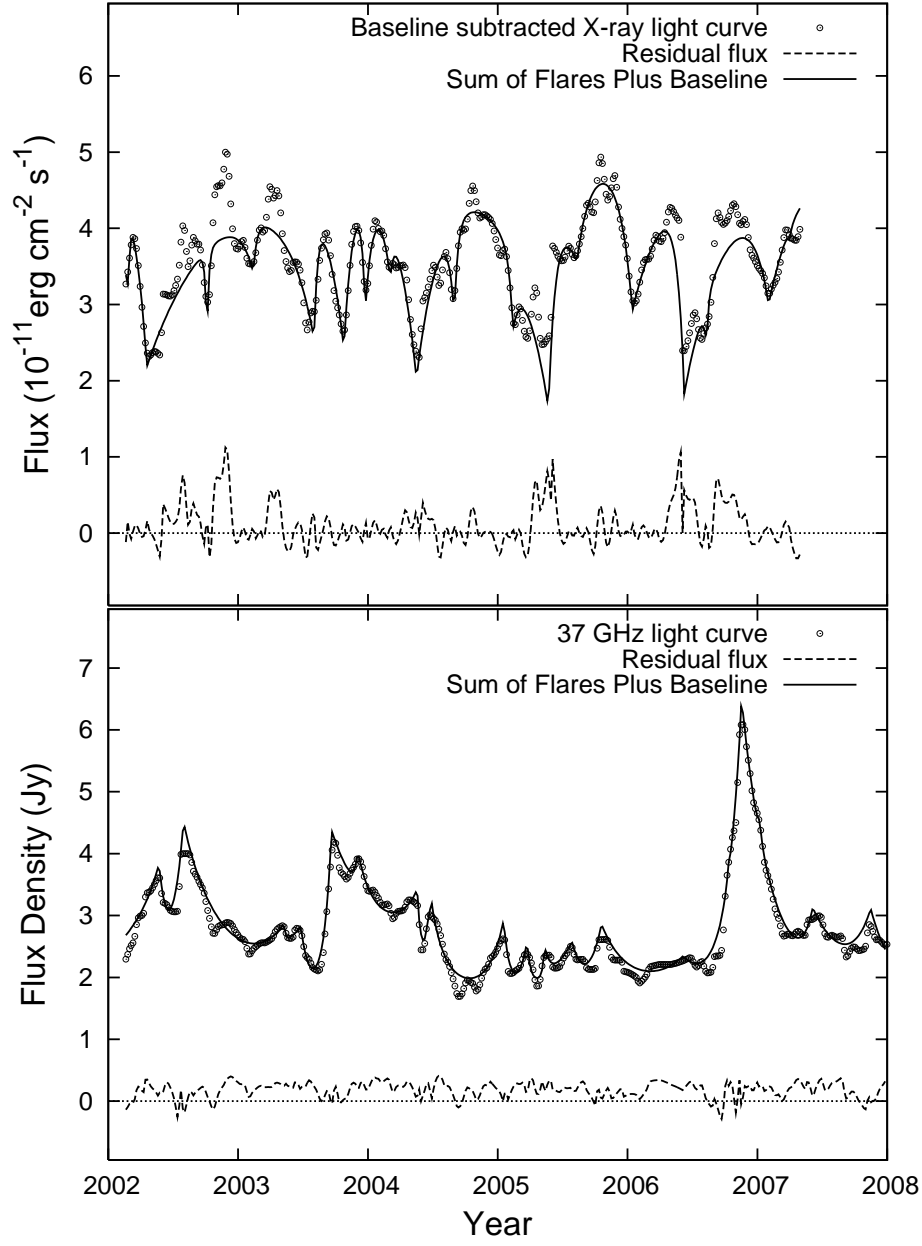}
\caption{X-ray and 37 GHz light curves. Curves correspond to summed flux after modeling the light curve as a superposition of many individual X-ray dips or radio flares and a baseline as shown in Figure~\ref{baseline1sig}.}
\label{xradmodelfit_1sig}
\end{figure}

There is a weak correlation between the X-ray spectral index and the X-ray flux, indicating that the spectrum becomes harder during a decrease in the production of X-rays (Fig \ref{baseline1sig}). This kind of ``pivoting" has been seen before \citep{ogl05,mar91}. The X-rays may be produced mainly by up-scattering of UV photons and not optical photons, consistent with the correlation found by \citet{ogl05} between the X-ray and UV variations in 3C 120. This could occur if the flux of optical photons reaching the corona is much smaller than that of UV photons. Such a scenario is likely if the corona is small such that the region where the UV photons are generated is much closer to the corona than the region where the majority of the optical photons are produced. Any disturbance propagating outwards in the accretion disk will cause a change in the UV flux (and a resultant nearly immediate change in the X-ray emission) followed by a similar change in the optical flux. The sign of the time delay will switch if the disturbance propagates inwards. The observed mean time delay between X-ray and optical variations may be due to such propagation time delays. If we adopt typical parameters such as $\sim$ 0.1 for accretion efficiency, L/L$_{ed}$ $\sim$ 0.3 \citet{ogl05}, and L$_{bol}\sim$ 2$\times$ $10^{45}$ ergs/sec \citep{woo02} and neglect General Relativistic effects, the region in the accretion disk where the emission peaks in the extreme UV/soft X-ray range ($\lambda=10$ nm) should be very close to the innermost stable orbit ($\sim$ 5 $r_g$, where $r_g$ is the gravitational radius of the black hole). The region where the emission peaks in the optical ($\lambda=600$ nm) lies at $\sim$ 1000$r_g$. For 3C 120, with a BH mass of $\sim$ 5$\times10^{7}$ M$_\sun$, 1000 $r_g$ is equivalent to $\sim$4 light days. We consider a model that includes both time delays and coupling of emission from the corona and disk.

\begin{figure}
\epsscale{0.8}
\plotone{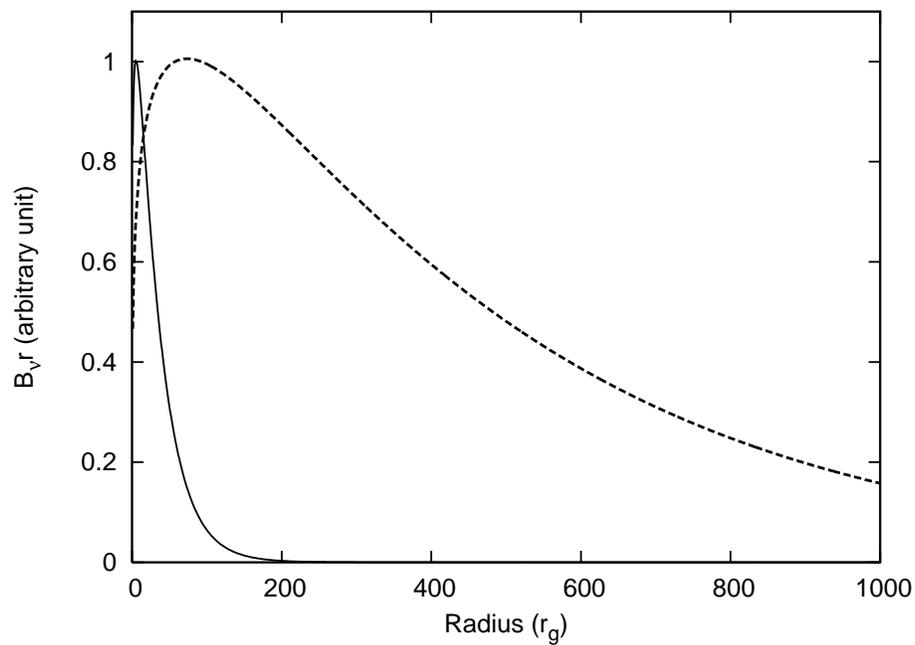}
\caption{Total intensity of radiation coming from different annuli of an accretion disk versus the annular radii. The solid and dashed curves correspond to UV and optical wavelengths, respectively.}
\label{theory0}
\end{figure}

We have performed a theoretical calculation to produce multi-wavelength light curves from a disk-corona system. We assume that (1) the temperature of the disk changes with its radial distance from the center according to T$\sim r^{-3/4}$ \citep{sak73}, (2) each annulus radiates as a perfect black-body, (3) the X-rays are produced in a spherical region (corona) close to the center of the disk by inverse-Compton scattering of the disk photons that reach the corona, and (4) lower energy radiation (UV-optical) is produced both by blackbody radiation in the disk and reheating of the disk by the X-rays from the corona. For simplicity, we assume that the energy distribution of electrons is uniform throughout the volume of the corona. The fraction of the UV-optical photons that reach the corona are up-scattered to X-rays and a part of the X-rays re-heat the disk to produce additional thermal photons. We treat the X-ray production in the corona as a reflection that increases the energy of the radiation. The computer code that we use for the model also follows the travel time of photons from different radii of the disk to the corona and \textit{vice versa}. The temporal nature of the emission from the corona and the part of the disk close to it are very similar owing to the small light travel time. Figure~\ref{theory0} shows the variation of the total UV and optical intensity emitted by an annulus at radius $r$ of thickness $\Delta r \propto r$, i.e., $B_{\nu}r$ versus $r$. This shows that the annulus that produces the largest amount of UV radiation (``region UV") lies $\sim$5 $r_g$ from the center, and for optical emission (`region Op") it is at $\sim$75 $r_g$. This distance is much smaller than what would be expected if most of the optical radiation were to come from radii near where the Planck curve $B_{\nu}$ peaks at R band. These latter radii lie $\sim$4 light-days (1000 $r_g$) from the center, while regions close to the center are at higher temperature, and therefore emit at a higher blackbody intensity at all wavelengths. Although $B_{\nu}$ increases monotonically with decreasing radii, $B_{\nu}r$ peaks at a radius $\sim$75 $r_g$ at R band.

\begin{figure}
\epsscale{0.8}
\plotone{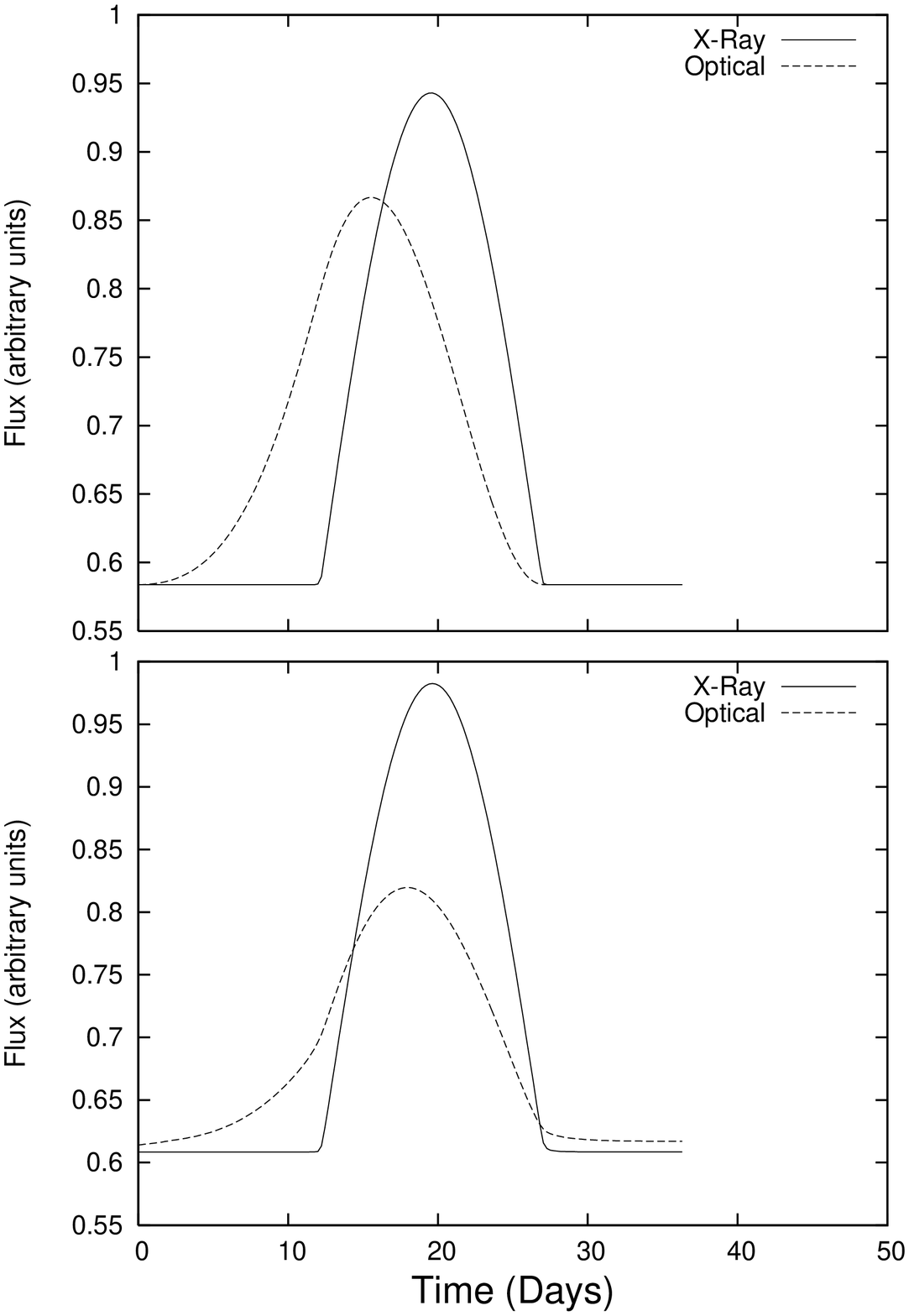}
\caption{Simulated X-ray and optical light curves from an accretion disk-corona system using our theoretical calculation. The disturbance is propagating inwards towards the BH. The top panel corresponds to zero feedback and the bottom panel shows the same after including feedback.}
\label{theory1}
\end{figure}

We use a computer code to model this disk-corona system and then introduce a disturbance in the temperature of the accretion disk that propagates at a speed $\sim$0.1$c$ from the center toward the outside or \textit{vice versa}. Due to the disturbance, the temperature at a given annulus varies with time according to a Gaussian profile. This causes a flare in the emission of the entire system at all wavelengths, although the flare starts and peaks at different times at different wavelengths. We produce these light curves at X-ray, UV, and optical bands, including delays from internal light-travel time. The nature of the flare and relation between flares at multiple wave bands depend on the speed and direction of propagation of the disturbance since emission from different annuli of the disk have different wavelength dependence. The top panel of Figure~\ref{theory1} shows the X-ray and optical light curves from the above calculation when the disturbance propagates toward smaller radii. As a result, the emission from the outer disk flares before that closer to the BH. Therefore, the optical variation leads that in the X-rays. In this simulation, the time delay between the peaks of the X-ray and optical light curves $\sim$4 days. 

\begin{figure}
\epsscale{0.8}
\plotone{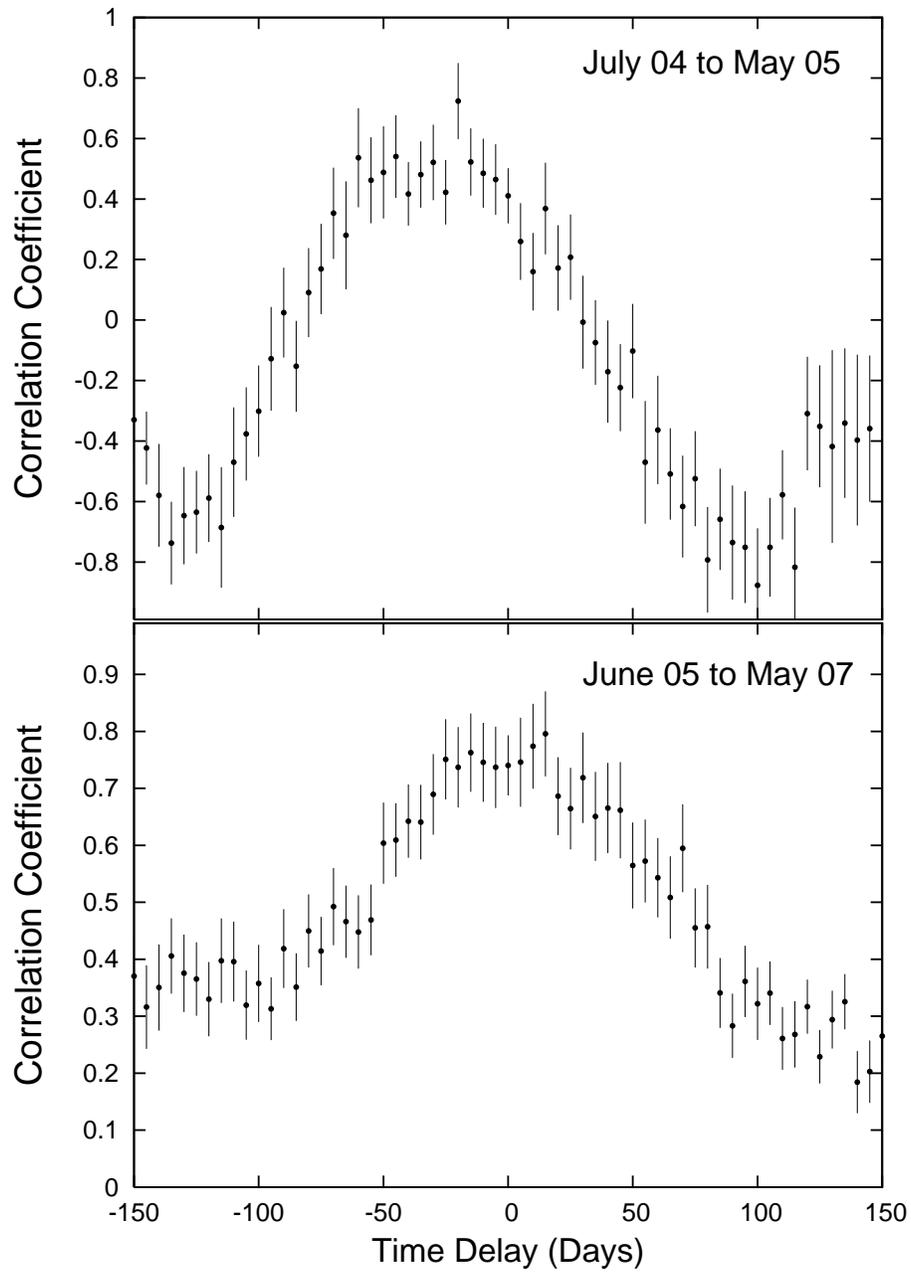}
\caption{Variation of the X-ray/optical correlation function across two intervals. The X-ray variations lead those in the optical by 25 days during the first interval and the two variations are almost simultaneous during the second interval.}
\label{xop_tw}
\end{figure}

In 3C 120, the time delay is centered on 0.5 day with an uncertainty of $\pm4$ days, hence the direction of the time delay cannot be specified with certainty. The correlation function in Figure \ref{xop_small} also shows a relatively broad peak centered on zero, with a similar uncertainty in the time delay. The cross-correlation function for longterm light curves sometimes show a broad peak due to changes in the direction and magnitude of the time delay over the timespan of the light curves. If the broad peak in these correlation functions is due to the above-mentioned reason, it may indicate a dichotomy in the speed of propagation of the disturbances and their directions, i.e., the uncertainty corresponds to an actual range of time delays. This also constrains the size of the corona. For example, if the corona were spread such that the regions Op and UV were at the same distance from the corona, then these regions would contribute equally to the flux of seed photons that are up-scattered to X-rays. In that case, the X-ray flares would be much broader than observed and no optical/X-ray time delay would be present. This is contrary to the observation of relatively sharp X-ray flares. In addition, although overall we find a short mean time delay of $0.5$ day (i.e., if the uncertainty does not correspond to an actual range of time delays but rather is consistent with zero), in some of the individual flares the time delay is longer ($\sim$25 days; see top panel of Figure~\ref{xop_tw}). Also, if optical and UV photons were up-scattered to soft and hard X-rays, respectively, a significant ``soft-hard" time delay would be expected. But from Figure~\ref{softhard}, it is clear that the soft-hard time lag cannot be more than 0.25 day, which is essentially unobservable with the longterm light curves. From the above discussion, we conclude that the corona is situated close to the center of the accretion disk, and its size is such that the flux of photons reaching the corona from region Op is negligible with respect to that from region UV. Based on the solid angle subtended, the coronal radius should be less than 40$r_g$ for the flux from region Op to be less than 10\% of that from region UV.

\begin{figure}
\epsscale{0.8}
\plotone{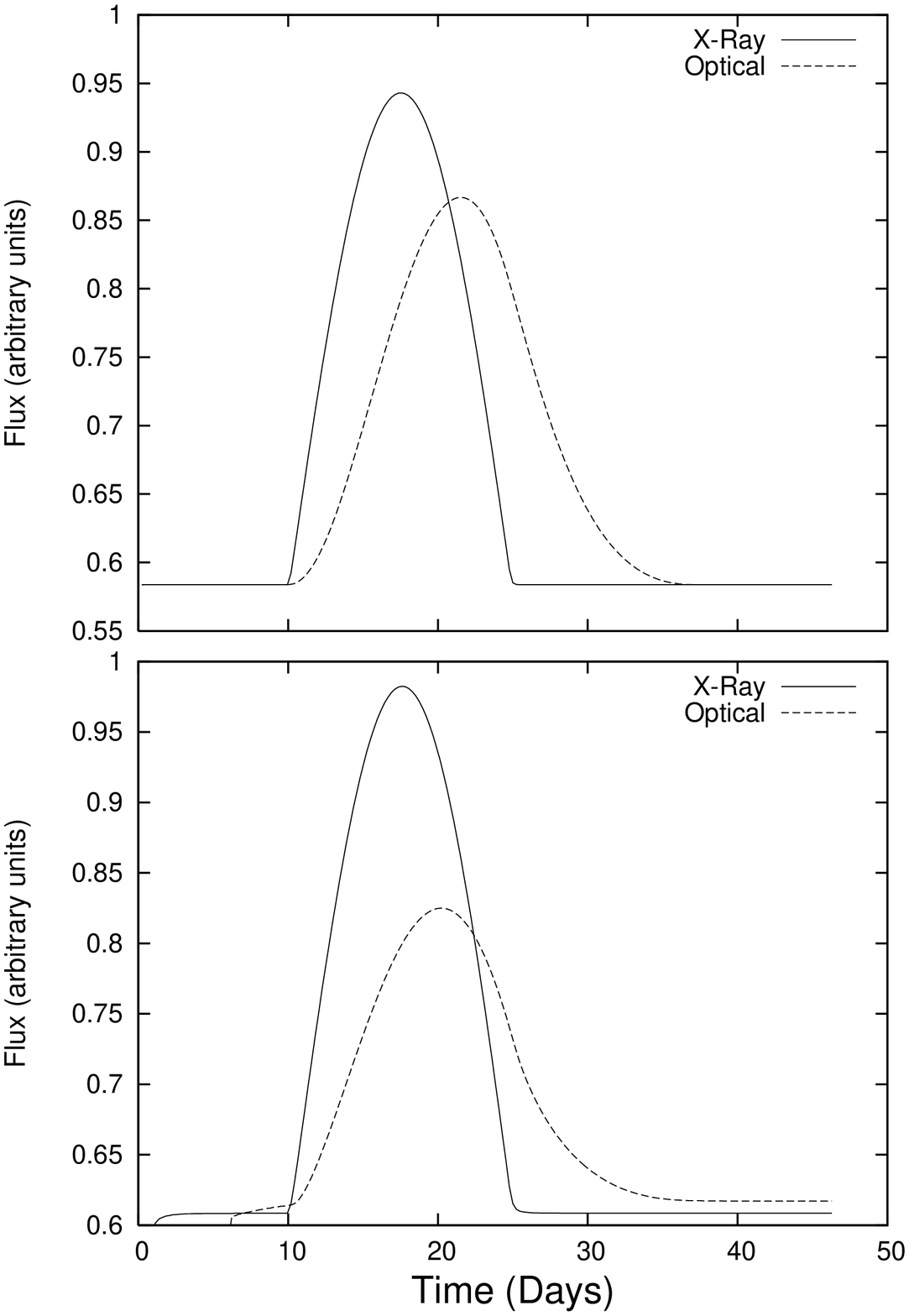}
\caption{Simulated X-ray and optical light curves from an accretion disk-corona system using our theoretical calculation. The disturbance is propagating outwards away from the BH. The top panel corresponds to zero feedback and the bottom panel shows the same after including feedback.}
\label{theory2}
\end{figure}

In Figure~\ref{theory1}, the two panels show results for different levels of feedback (the fraction of the X-rays produced in the corona that reflect toward and re-heat the disk). The top panel has feedback fraction equal to zero, i.e., none of the X-rays re-heat the disk. We can see from the figure that, as feedback is introduced, the resulting time delay becomes smaller and less precise. Figure~\ref{theory2} shows the light curves when the disturbance is propagating outwards from the BH. In this case, as expected, the X-ray variations lead those in the optical (by similar amount as in Figure~\ref{theory1}) for the same propagation speed. The panels show, as above, that including feedback makes the time delay shorter. Hence, feedback from the corona may also contribute to the range of time delays that we see in the data.

In another possible scenario, the X-ray variability is caused by intrinsic changes in the hot electrons in the corona and the UV-optical changes are due to feedback, i.e., there is no intrinsic variability in the accretion disk. In this case, the X-ray/optical time delay will solely be due to light travel time from the corona to the accretion disk. This will produce time lags of a small fraction of a day \citep{kaz01}, which is too small to observe with the sampling of the data used in this work. The observed essentially zero time lag between the long-term X-ray and optical light curves ($0.5\pm4$ days) and smaller variability amplitude in the short-term optical light curve than that in the X-rays are consistent with the above scenario. In fact the steepening of the X-ray spectrum and decrease in the mean X-ray flux level after 2006.0 (see Fig. \ref{baseline1sig}) suggest a long-term steepening of the energy distribution of electrons in the corona. But the longer X-ray/optical time delay over a significant portion of the dataset (Figure~\ref{xop_tw}) and comparable X-ray/optical B-band variability amplitude in the long-term light curves \citep{dor09} with B-band variations sometimes leading, indicate that propagation of disturbances in the accretion disk must produce at least part of the X-ray/optical variability in this radio galaxy.

\begin{figure}
\epsscale{1.0}
\plotone{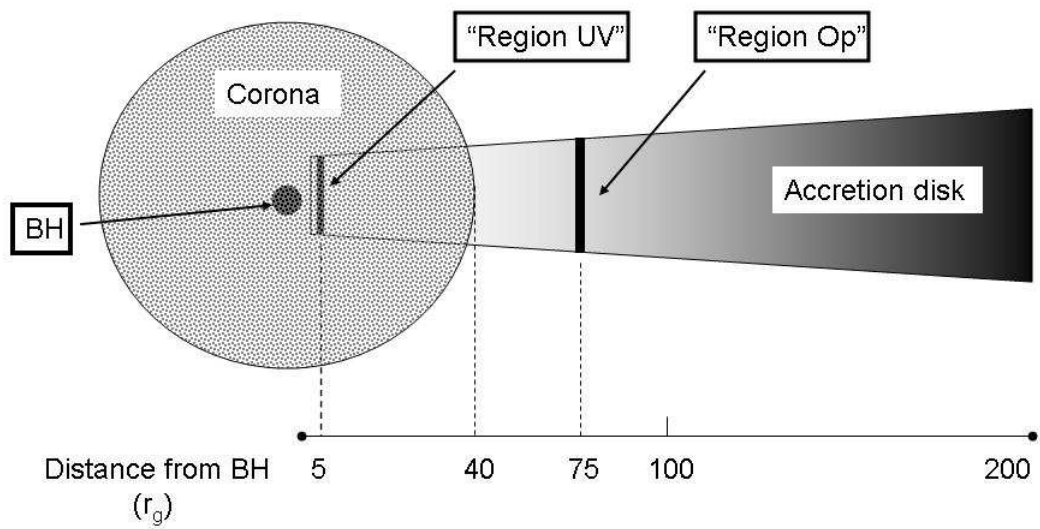}
\caption{Sketch of the accretion disk-corona system as derived in this paper.}
\label{cartoon}
\end{figure}

Figure~\ref{cartoon} shows a sketch of the accretion disk-corona system as derived in this paper. The black filled circle is the position of the BH, the temperature of the accretion disk is shown by gray scale with lighter color meaning higher temperature, and the larger circular area filled with dots is the corona. ``Region UV" and ``Region Op" are shown as thick solid lines on the accretion disk. The radius of the corona and the distance of the relevant emission regions from the BH are shown in units of $r_g$.

From the above model and the correlation results, we conclude that most, if not all, of the optical emission in 3C 120 is produced in the accretion disk. The X-rays are produced by scattering of (mostly) UV radiation in the corona. Optical/UV emission due to re-heating of the disk by the X-rays is also a possible ongoing mechanism that may cause the spread of the time delay. The presence of sharp X-ray flares and the X-ray/optical time delay of weeks in some individual flares indicate that the corona must lie sufficiently close to the BH that the flux of seed photons reaching the corona is dominated by UV light. 

The light curve during the dip at X-ray and optical wave bands between 2006 October and 2007 April, displayed in Figure~\ref{xop_commondips}, is very well-sampled. It is evident that the optical flux starts to decrease after $\sim$MJD $4025$, $\sim 40$ days earlier than the decrease starts in the X-rays. The minimum in optical occurs $\sim$40 days earlier than that in the X-rays as well. This indicates that the decrease was caused by a disturbance propagating from the outer radii of the accretion disk toward the BH. If this disturbance is a thermal fluctuation propagating inward, then it should have an effective speed $\lesssim$ $0.01c$ to cause a time delay of $\sim$40 days. This is one order of magnitude higher than the sound speed for a gas pressure dominated disk of temperature $\leq$ $10^5$ K. Therefore, the above disturbance cannot be transmitted by sound waves \citep{kro91} unless the relevant regions of the disk are dominated by radiation pressure, which cannot be ruled out in 3C 120 given its high accretion rate, nearly the Eddington value (0.3 $L_E$). On the other hand, this scenario is consistent with the model proposed by \citet{kin04}, in which large scale alignment of poloidal magnetic field in the inner accretion disk from random fluctuations causes the decrease in the X-ray flux. Such alignment occurs at a timescale $2^{R/H}k(R^3/GM)^{1/2}$, where $R/H$ is the radius to thickness ratio of the disk, $k$ is a constant $\sim$10 \citep{tou92,sto96}, and $(R^3/GM)^{1/2}$ is the disk dynamical timescale at radius $R$. This alignment timescale, for a $10 M_{\sun}$ BH, is few seconds, which translates to $\sim$50 days for a $\sim5\times10^7 M_{\sun}$ BH in 3C 120. Model light curves of \citet{kin04} also contain short timescale, small amplitude fluctuations on top of the big flares and dips caused by small-scale alignment of the poloidal magnetic field, similar to that observed in the light curves of 3C 120 presented in this paper.

\section{Summary and Conclusions}
This paper presents well-sampled, 5-yr-long light curves of 3C 120 between 2002 and 2007 at X-ray, optical, and radio wavebands, as well as monthly images obtained with the VLBA at 43 GHz. We have calculated the broad-band X-ray power spectral density (PSD) of 3C 120. The X-ray and radio light curves were decomposed into individual dips and flares, respectively, and the properties of the contemporaneous dips and flares were compared. That comparison, as well as, the X-ray/radio correlation analysis were used to find a connection between events in the jet and the accretion disk. We have also cross-correlated the X-ray and optical light curves and have determined the significance of the correlations with simulated light curves based on the PSD. Using a numerical model, we have produced X-ray and optical light curves from the accretion disk-corona system. From the comparison of the temporal properties of the simulated light curves to that of the observed variability, we infer the physical size of the corona and the distances of the emitting regions from the central BH.  

Our main conclusions are as follows: \\
(1) The X-ray PSD of 3C 120 is best fit by a bending power-law model where there is a smooth change in the slope above a break frequency. The best fit value of break frequency for 3C 120 is $10^{-5.05}$ Hz, which agrees very well with the relation between break timescale, BH mass and accretion rate obtained by \citet{mch06} spanning a range of BH mass from 10 $M_{\sun}$ to $10^9$ $M_{\sun}$. This indicates that the accretion process in 3C 120 is similar to that of the BHXRBs.\\
(2) The X-ray and 37 GHz variations are anti-correlated with the X-ray leading the radio by $\sim$120 days. The anti-correlation remains even if the large X-ray dips and radio flares after 2006 are excluded.\\
(3) Almost all X-ray dips are followed by the ejection of a new knot in the VLBA images. This and the anti-correlation mentioned in (2) imply that decrease in the X-ray production is linked with increased speed in the jet flow, causing a shock front to form and move downstream. This property of 3C 120 is also similar to the galactic black hole systems where transitions to high-soft X-ray states are associated with the emergence of very bright features that proceed to propagate down the radio jet.\\
(4) We derive a distance $\sim$0.5 pc from the corona (where the X-rays are produced) to the VLBA 43 GHz core region using the average time delay between the start of the X-ray dips and the time of ``ejection" of the corresponding superluminal knots. \\
(5) The X-ray and optical variations in 3C 120 are very strongly correlated (see also \citet{mar09,dor09}). This correlation, absence of significant optical polarization, and anti-correlation of X-ray and 37 GHz variations (the latter of which are produced in the jet), together imply that the optical emission is blackbody radiation from the accretion disk. \\
(6) The X-rays are produced by scattering of (mostly) UV radiation in the corona. Comparison of simulated light curves from a disk-corona system and the observed variation imply that the corona must lie sufficiently close to the BH that the flux of seed photons reaching the corona is dominated by UV light.

\section{Acknowledgments}
We thank the anonymous referee for many useful comments which helped to improve the quality of the manuscript. We would also like to thank P. Uttley for many valuable discussions. The research at Boston University was funded in part by the National Science Foundation (NSF) through grant AST-0406865 and by NASA through several RXTE Guest Investigator Program grants, most recently NNG05GM33G, NNG05GM64G, and most recently Astrophysical Data Analysis Program grant NNX08AJ64G. The Mets\"ahovi team acknowledges the support from the Academy of Finland. The University of Michigan Radio Astronomy Observatory was supported by funds from the NSF, NASA, and the University of Michigan. The VLBA is an instrument of the National Radio Astronomy Observatory, a facility of the National Science Foundation operated under cooperative agreement by Associated Universities, Inc. The Liverpool Telescope is operated on the island of La Palma by Liverpool John Moores University in the Spanish Observatorio del Roque de los Muchachos of the Instituto de Astrofisica de Canarias, with financial support from the UK Science and Technology Facilities Council. MG thanks the US National Science Foundation for support through grants AST 03-07912 and AST 08-03883. HP, JT, and ZS thank the Laboratory for the Optical Astronomy of the Chinese Academy of Sciences for partial support and the National Natural Science Foundation of China for support through grants 10633010 and 10625314.


\clearpage

\clearpage

\clearpage
\begin{table}
\begin{center}
\caption{Parameters of the Light Curves.\label{data}}
\begin{tabular}{ccccccc}\\
\tableline
  &Data set & Start & End & T (days)\tablenotemark{1} & $\Delta$T (days) \tablenotemark{2}  & N$_{points}$ \\
\tableline
        & Longlook & 2002 December 13 & 2002 December 14 & 1.5 & 0.01& 150 \\
 X-ray  & Medium & 2006 November & 2007 January  &60.0  &0.25  & 240\\
        & Monitor & 2002 March & 2007 May &1910.0 &15.0 & 1050\\ 
\tableline
        
 Optical & Monitor &2004 August & 2008 January  &1250.0 &  -  & 154 \\
        
\tableline
 Radio & Monitor &2002 March & 2008 January  &2167.0  &  -    & 329\\
         
\tableline
\end{tabular}
\tablenotetext{1}{Total length of light curves}
\tablenotetext{2}{Bin size}
\end{center}
\end{table}

\begin{table}
\begin{center}
\caption{Time, Area and Width of the X-ray Dips and 37 GHz Flares, and Times of Superluminal Ejections.\label{ejecflare}}
\begin{tabular}{ccccccccc}\\
\tableline\tableline
\multicolumn{4}{c}{Parameters of X-ray Dips}&Ejection Times & Knot ID &\multicolumn{3}{c}{Parameters of 37 GHz Flares}\\ 
Time (start) & Time (min.)  & Area \tablenotemark{1}& Width\tablenotemark{2}&$T_0$ && Time (peak)  & Area\tablenotemark{3}& Width\tablenotemark{2} \\
\tableline

 $-$\tablenotemark{4}& 2002.15 & $-$\tablenotemark{4}&$-$\tablenotemark{4}& 2002.23 $\pm$ 0.03&02A&2002.39  & 195 &  95\\
  2002.19 &    2002.30 &  657   &  120.  &  2002.65  $\pm$ 0.04& 02B &   2002.58  & 404 &  75 \\
  2002.75 &    2002.76 &  12    &    5.  &                     &     &            &     &     \\
  2003.02 &    2003.12 &   29   &   22.5 &  2003.35  $\pm$ 0.15& 03A &   2003.35  &  17 &  17 \\
  2003.32 &    2003.58 &   172  &   52.5 &  2003.67  $\pm$ 0.02& 03B &   2003.72  & 441 &  85 \\
  2003.66 &    2003.82 &   184  &   40.  &  2003.81  $\pm$ 0.03& 03C &   2003.92  &  24 &  20 \\
  2003.95 &    2003.98 &   44   &   15.  &  2003.98  $\pm$ 0.03& 03D &            &     &     \\
  2004.12 &    2004.17 &   20   &   15.  &  2004.16  $\pm$ 0.05& 04A &            &     &     \\
  2004.21 &    2004.37 &   364  &   65.  &  2004.37  $\pm$ 0.03& 04B &   2004.38  & 126 &  52 \\
          &            &        &        &                     &     &   2004.49  &  60 &  25 \\
  2004.62 &   2004.66  &   30   &   12.5 &  2004.82  $\pm$ 0.05& 04C &   2005.05  &  61 &  27 \\
  2005.09 &    2005.12 &   33   &   17.5 &  2005.14  $\pm$ 0.03& 05A &   2005.23  &  21 &  15 \\
  2005.19 &    2005.39 &  351   &   65.  &  2005.34  $\pm$ 0.02& 05B &   2005.36  &  10 &  10 \\ 
	  & 	       & 	&        &		       &     &   2005.57  &  92 &  57 \\ 	
	  & 	       & 	&        &		       &     &   2005.80  & 127 &  57 \\ 	
  2005.94 &   2006.04  &  287   &   70.  &  2006.00  $\pm$ 0.03& 06A &   2006.43  & 126 &  105\\
  2006.38 &   2006.44  &   382  &   72.5 &  2006.72  $\pm$ 0.05& 06B &   2006.88  & 598 &  62 \\
  2006.96 &   2007.09  &  193   &   62.5 &  2007.05  $\pm$ 0.02& 07A &   2007.43  & 132 &  72 \\
          &            &        &        &  $-$\tablenotemark{4}  &     &   2007.87  & 122 &  60 \\
\tableline 
\end{tabular}
\tablenotetext{1}{Equivalent width of dips. Units: $10^{-6}$ erg cm$^{-2}$}
\tablenotetext{2}{Units: days}
\tablenotetext{3}{Area under the flare light curves. Units: $10^{-6}$ erg cm$^{-2}$}
\tablenotetext{4}{Insufficient data}
\end{center}
\end{table}

\end{document}